\documentclass[10pt]{iopart}

\usepackage{graphicx}
\begin{document}

 \title{Octupole deformation in neutron-rich actinides and superheavy 
nuclei and the role of nodal structure of single-particle wavefunctions 
in extremely deformed structures of light nuclei.}

\author{A.\ V.\ Afanasjev$^{1,3}$, H.\ Abusara$^{2}$, S.\ E.\ Agbemava$^{1}$}
\address{$^{1}$Department of Physics and Astronomy, Mississippi
State University, MS 39762, USA}

\address{$^2$Physics Department, Birzeit University, Birzeit, Palestine}

\ead{$^3$Anatoli.Afanasjev@gmail.com}

\vspace{10pt}

\begin{abstract}
  Octupole deformed shapes in neutron-rich actinides and superheavy 
nuclei as well as extremely deformed shapes of the $N\sim Z$ light
nuclei have been investigated within the framework of covariant 
density functional theory. We confirmed the presence of new region of 
octupole deformation in neutron-rich actinides with the center 
around $Z\sim 96, N\sim 196$ but our calculations do not predict
octupole deformation in the ground states of superheavy $Z\geq 108$
nuclei. As exemplified by the study of $^{36}$Ar, the nodal structure 
of the wavefunction of occupied single-particle orbitals in extremely 
deformed structures allows to understand the formation of the $\alpha$-clusters 
in very light nuclei, the suppression of the $\alpha$-clusterization 
with the increase of mass number, the formation of ellipsoidal 
mean-field  type structures and nuclear molecules.
\end{abstract}


%
%
%
%
\ioptwocol
%
\begin{figure*}[ht]
\includegraphics[angle=-90,width=8.5cm]{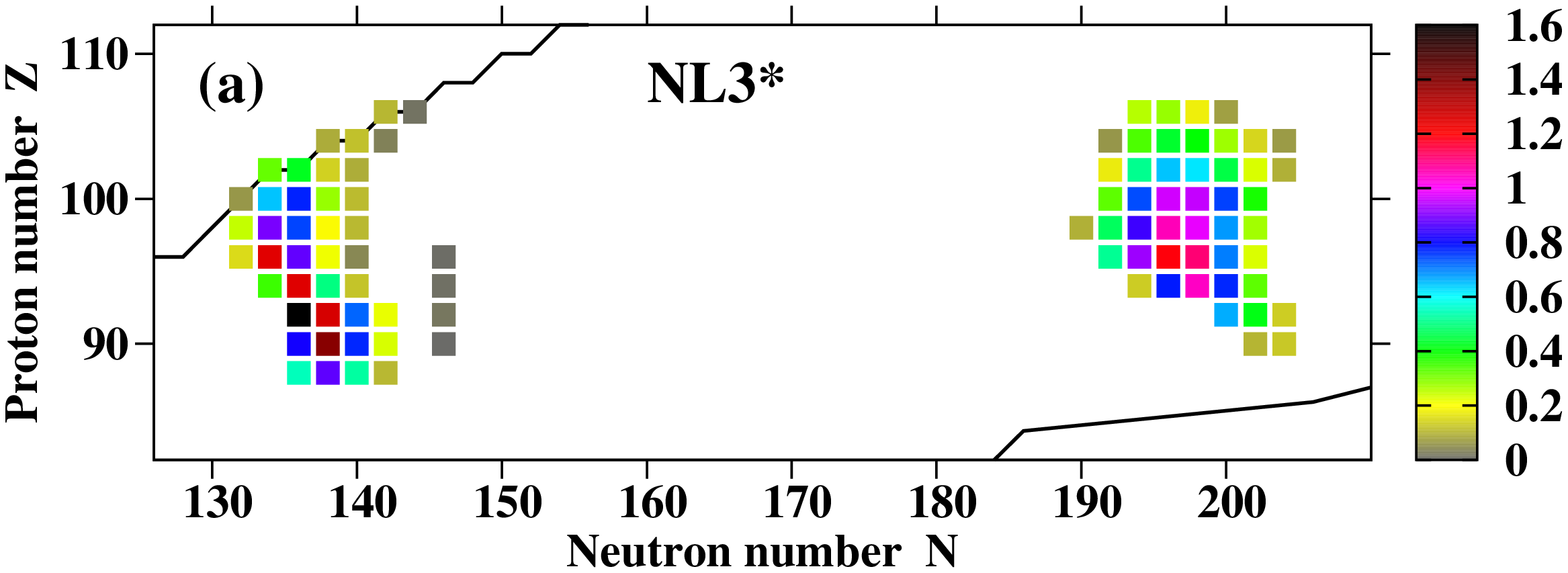}
\includegraphics[angle=-90,width=8.5cm]{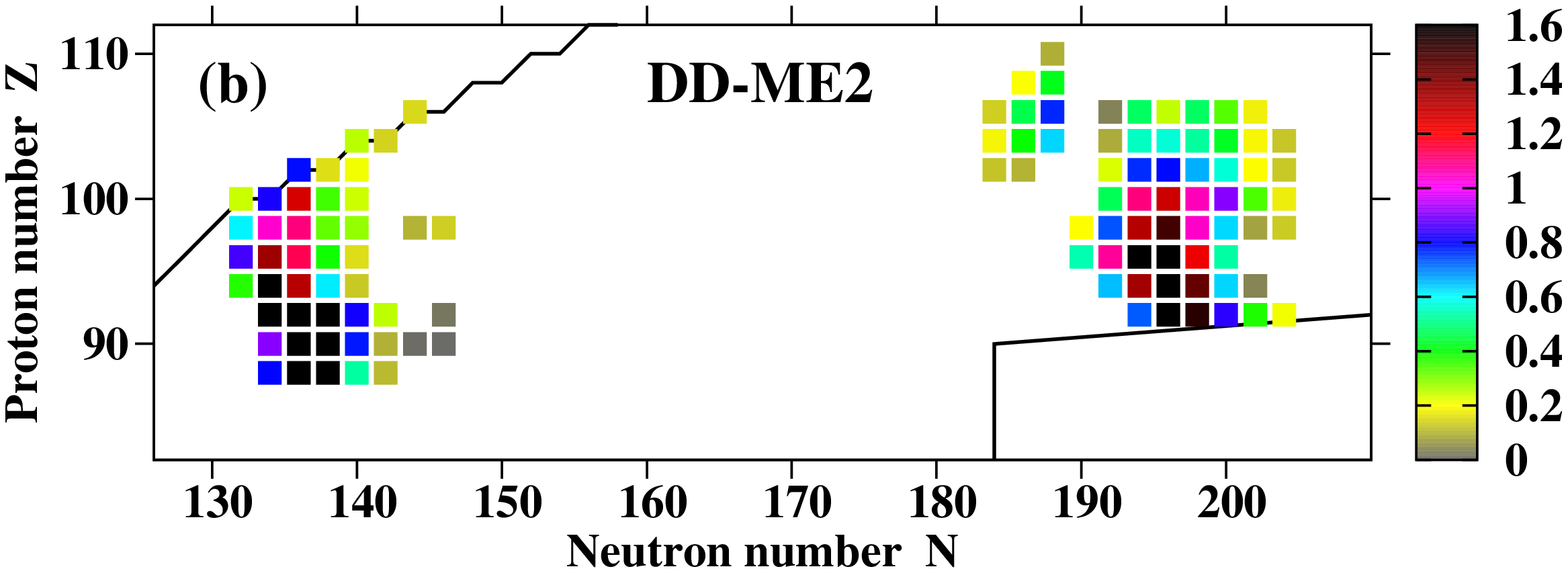}
\includegraphics[angle=-90,width=8.5cm]{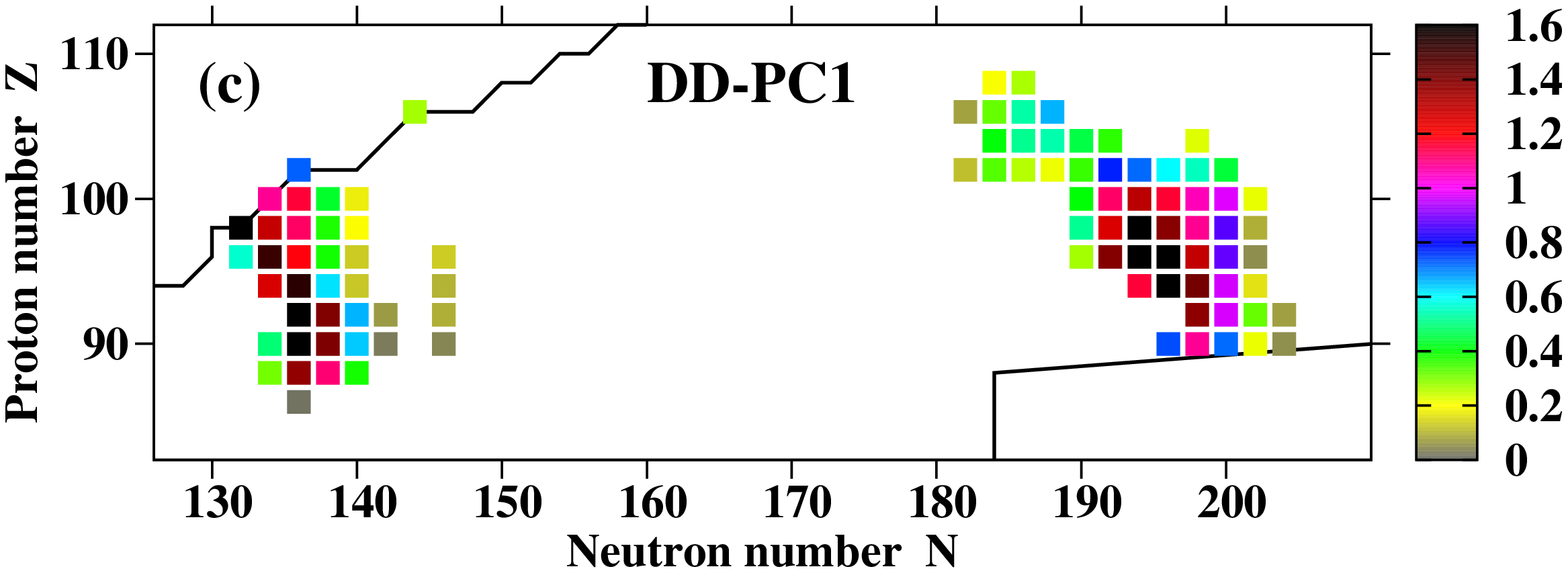}
\includegraphics[angle=-90,width=8.5cm]{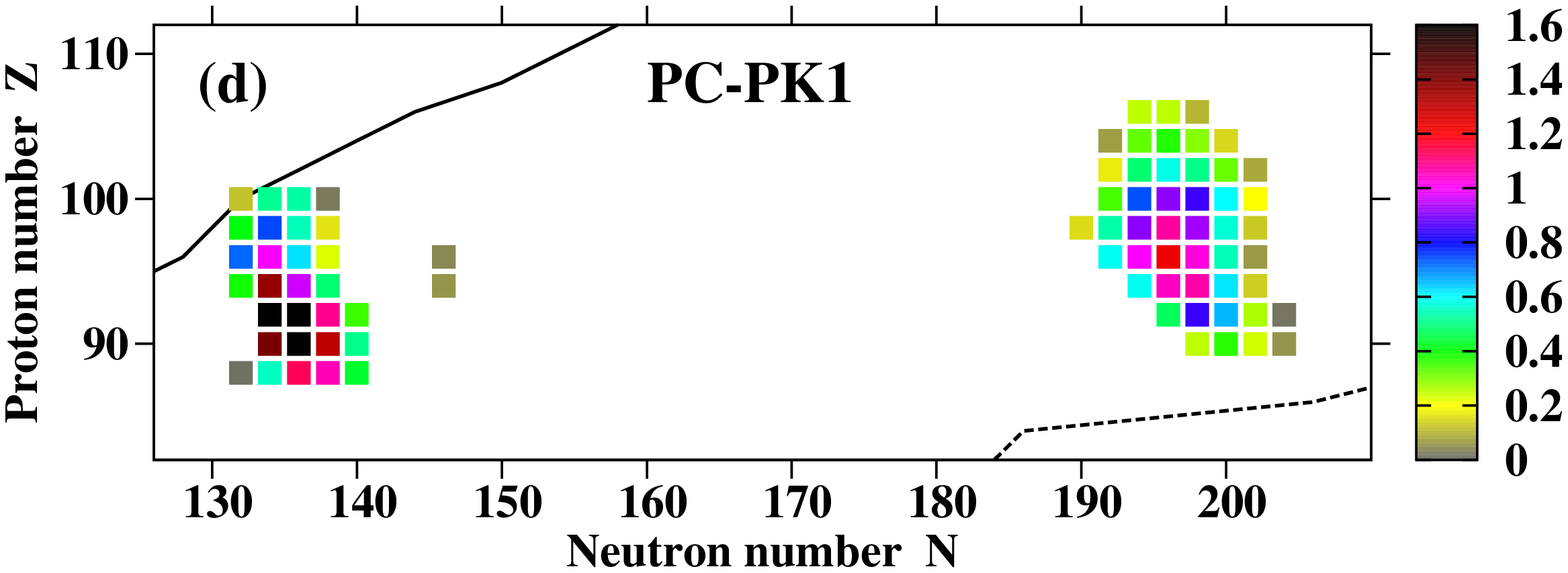}
  \caption{Octupole deformed nuclei in the part of nuclear chart
           under study for indicated covariant energy density
           functionals. Only nuclei with non-vanishing $\Delta E^{oct}$ 
           are shown by squares; the colors of the squares represent the values of
           $|\Delta E^{oct}|$  (in MeV) (see colormap). The two-proton and
           two-neutron drip lines are displayed by solid black
           lines. From Ref.\ \cite{AbgemA.17}.
            }
\label{fig-global-oct-CDFT}
\end{figure*}

\section{Introduction}

  The concepts of nuclear shape and shape coexistence are 
the centerpieces of low energy nuclear physics \cite{NilRag-book}. 
These shapes are connected with the symmetry breaking of 
the nuclear mean field and manifest themselves in different 
forms. Breaking of spherical symmetry leads to deformed
shapes, the simplest ones are axial quadrupole deformed shapes. 
However, the change of their elongation leads to different 
classes of nuclear shapes such as normal-, super-, hyper- and
megadeformed ones. Next step is breaking of the symmetry
of the mean field in the plane perpendicular to the axis
of symmetry. This leads to reflection asymmetric (octupole
deformed) shapes. The density functional theory \cite{BHP.03,VALR.05} 
provides a natural framework for the description of different
classes of nuclear shapes across whole nuclear chart. This manuscript 
presents recent results obtained in the studies of nuclear shapes 
within the framework of covariant density functional theory (CDFT) 
\cite{VALR.05}. It is focused on two issues discussed 
below and covers two extreme ends of the nuclear chart. 

  First issue is the role of octupole deformation in the ground 
states of neutron-rich actinides and superheavy nuclei. The global 
investigation of Ref.\ \cite{AAR.16} performed with the DD-PC1 \cite{DD-PC1} 
and NL3* \cite{NL3*} covariant energy density functionals (CEDFs) found the presence 
of the island of octupole deformation in the region with center around 
$Z\sim 96, N\sim 196$. In order to estimate theoretical uncertainties 
in model predictions, we performed additional studies with the DD-ME2 \cite{DD-ME2} 
and PC-PK1 \cite{PC-PK1} CEDFs. This study covers not only the above mentioned region 
but also extends to superheavy nuclei for which  the calculations
have been performed with all four functionals. Note that the octupole 
deformation in the ground states of superheavy nuclei has not been studied  
in the CDFT framework before our investigation.

 The second issue is the role of the single-particle degrees of freedom 
in the formation of extremely deformed shapes of rotating nuclei and in the 
transition from ellipsoidal mean field type configurations towards nuclear 
molecules. A systematic investigation of extremely deformed structures at 
high spin has been performed in Refs.\ \cite{RA.16,AR.17} for the 
$N \approx Z$ nuclei with $Z=14-24$. These studies show that particle-hole 
excitations within the same nucleus lead to the formation of different 
nuclear shapes starting from spherical ones via normal-deformed to super-, 
hyper- and megadeformed ones. Among these extremely deformed shapes there 
are the examples of ellipsoidal mean-field type structures, nuclear molecules 
and clustered configurations. Thus, it is important to understand what role 
single-particle states (and, in particular, the nodal structure of their
wavefunctions) are playing in the formation of such structures. To 
our knowledge, this aspect of the nuclear many-body problem has not been 
studied so far.

   The paper is organized as follows. Section \ref{Octupole} describes
the main results obtained in the study of octupole deformation in the ground
states of neutron-rich actinides and superheavy nuclei.  Sec.\ \ref{36Ar} 
is devoted to the discussion of the role of the single-particle degrees 
of freedom in clusterization and in the formation of extremely deformed 
structures and nuclear molecules; this is done on the example of $^{36}$Ar. 
Finally, Sec.\ \ref{conclusions} summarizes the results of our work.

\section{Octupole deformation in neutron-rich actinides
         and superheavy nuclei}
\label{Octupole}

  The calculations have been performed in the Relativistic-Hartree-Bogoliubov 
(RHB) approach using parallel computer code RHB-OCT developed in Ref.\ 
\cite{AAR.16}. In the calculations, the constraints on quadrupole and
octupole moments are employed. In order to avoid the uncertainties connected 
with the definition of the size of the pairing window \cite{KALR.10}, the separable 
form of the finite range Gogny pairing interaction introduced by Tian et al 
\cite{TMR.09} is used in the calculations.

  The effect of octupole deformation is characterized by the 
quantity $\Delta E_{oct}$ defined as
\begin{equation}
\Delta E_{oct} = E^{oct}(\beta_2, \beta_3) - E^{quad}(\beta'_2,\beta'_3=0)
\end{equation}
where $E^{oct}(\beta_2, \beta_3)$ and $E^{quad}(\beta'_2, \beta'_3=0)$
are the binding energies of the nucleus in two local minima of
potential energy surface (PES); the first minimum corresponds to octupole
deformed shapes and second one to the shapes with no octupole
deformation. The quantity  $|\Delta E_{oct}|$ represents the gain of
binding due to octupole deformation. It is also an indicator of
the stability of the octupole deformed shapes. Large $|\Delta E_{oct}|$
values are typical for well pronounced octupole minima in the
PES; for such systems the stabilization 
of static octupole deformation
is likely. On the contrary, small $|\Delta E_{oct}|$ values are characteristic
for soft (in octupole direction) PES typical for octupole vibrations.

\begin{figure}[htb]
\includegraphics[angle=0,width=8.75cm]{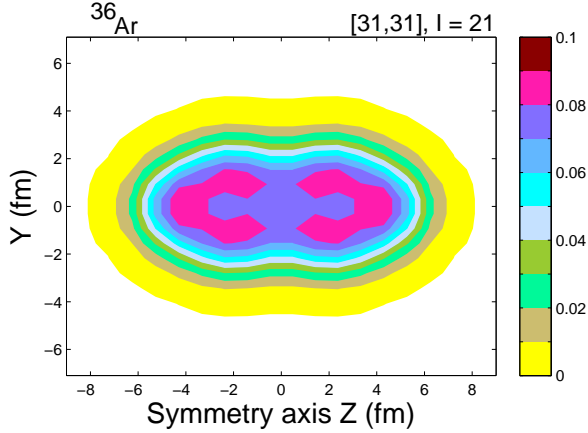}
\caption{Total neutron density [in fm$^{-3}$] of the megadeformed
[31,31] configuration in $^{36}$Ar obtained in the CRMF 
calculations with the NL3* functional. }
\label{Total-densities}
\end{figure}

 The RHB results for octupole deformed nuclei are summarized in Fig.
\ref{fig-global-oct-CDFT}. The present investigation confirms
the predictions of Ref.\ \cite{AAR.16} about the existence of the region 
of octupole deformation centered around $Z\sim 96, N\sim 196$ obtained 
with the DD-PC1 and NL3* functionals. Most of the CEDFs predict the size of 
this region in the $(Z,N)$ plane larger than the one of the experimentally
known region at $Z\sim 92, 
N\sim 136$. On the other hand, the impact of octupole deformation on the 
binding energies of the nuclei in these two regions are comparable. The 
search for octupole deformation in the ground states of even-even superheavy 
$Z=108-126$ nuclei with neutron numbers from the two-proton drip line up 
to neutron number $N = 210$ has been performed in the CDFT framework 
for the first time. 
With the  exception of two $Z=108$ (two $Z=108$ and one 
$Z=110$) octupole deformed nuclei in the calculations with CEDF DD-PC1 
(DD-ME2), no octupole deformed shapes in the ground states 
of these nuclei have been found.

  It is important to compare the CDFT predictions with the ones 
obtained in non-relativistic theories. Similar region of octupole 
deformation is predicted also in Skyrme DFT \cite{ELLMR.12} 
and microscopic+macrosscopic (mic+mac) \cite{MNMS.95} calculations. 
However, it is centered at $Z\sim 100, N\sim 190$ in the Skyrme 
DFT calculations and at $Z\sim 100, N\sim 184$ in mic+mac calculations. 
The existing Gogny DFT calculations \cite{WE.12} do not extend below 
$Z=98$ and beyond $N=190$; however, the trends seen in these 
calculations do not suggest the existence of the region of octupole 
deformation in very neutron rich actinides. The predictions for 
octupole deformation in the ground states of superheavy $Z\geq 108$ 
nuclei differ drastically. Both CDFT and Skyrme DFT do not 
predict octupole deformation in these nuclei. On the contrary,
large region of octupole deformation is predicted in superheavy
nuclei in the mic+mac and Gogny DFT calculations. These differences 
in the location of the islands of octupole deformed nuclei are due 
to  the differences in the underlying single-particle structure 
which exist among the models in actinides and superheavy nuclei
\cite{DABRS.15,BRRMG.99,AANR.15}.

  Note that the accounting of octupole deformation in the ground 
states of the $Z\sim 98, N\sim 196$ nuclei is essential for the modeling 
of fission recycling in neutron star mergers \cite{GBJ.11,JBPGJ.15} 
since the gain in binding energy of the ground states due to octupole 
deformation will increase the fission barrier heights as compared with 
the case when octupole deformation is neglected. These changes in binding
energy of the ground states and fission barriers affect the r-process
\cite{GBJ.11,JBPGJ.15}. It is also necessary to recognize that the 
present results are restricted to the mean field level. The methods beyond 
mean field such as quadrupole-octupole collective Hamiltonian \cite{XTLNV.17} or 
generator coordinate method including octupole deformation \cite{RB.11} have 
to be employed to define excitation spectra and transition rates of these
nuclei.

\section{The role of single-particle degrees of freedom
         in clusterization and nuclear molecules: an example 
         of megadeformed [31,31] configuration in $^{36}$Ar.}
\label{36Ar}

\begin{figure*}[htb]
\includegraphics[angle=0,width=5.45cm]{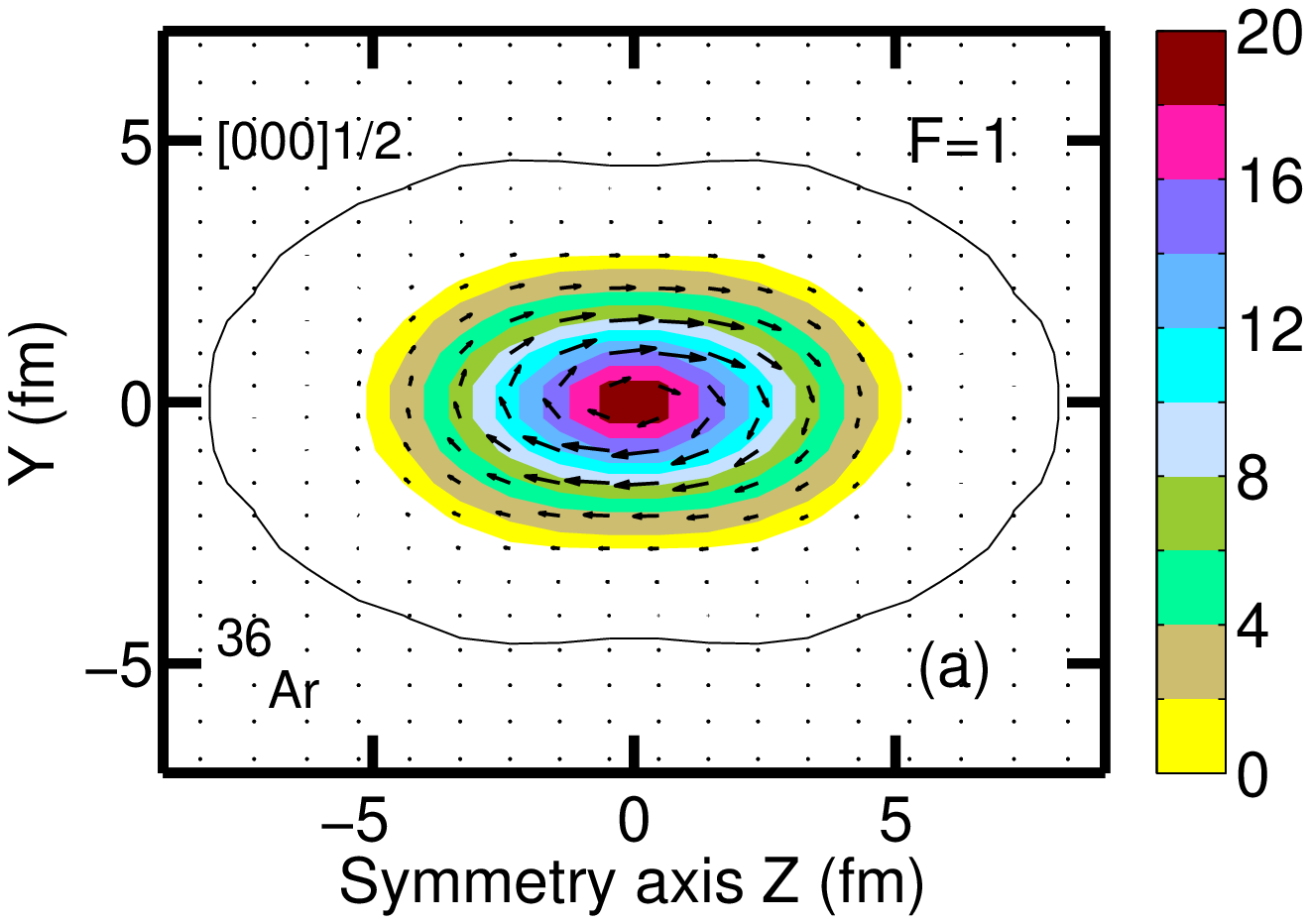}
\includegraphics[angle=0,width=5.45cm]{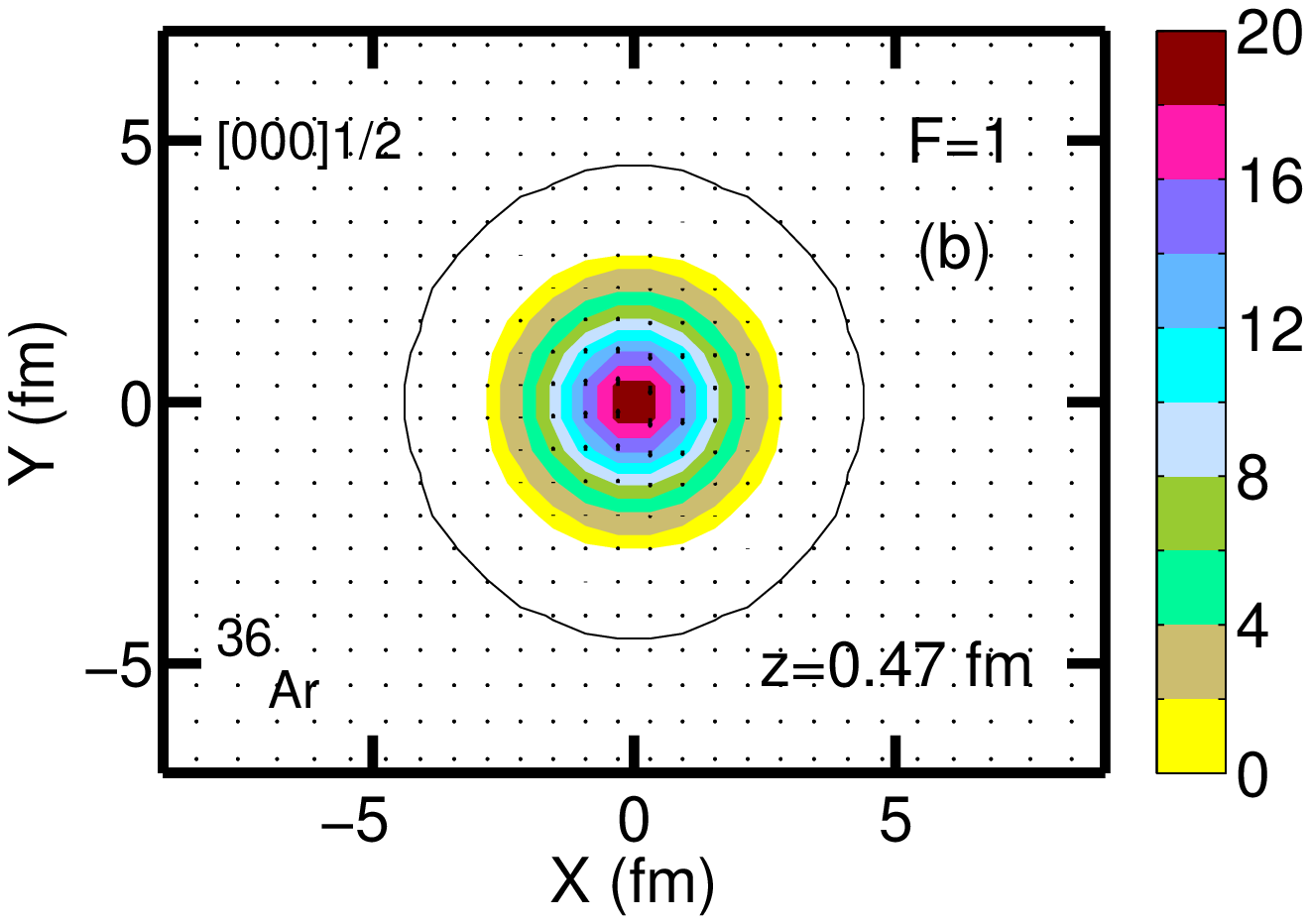}
\includegraphics[angle=0,width=5.45cm]{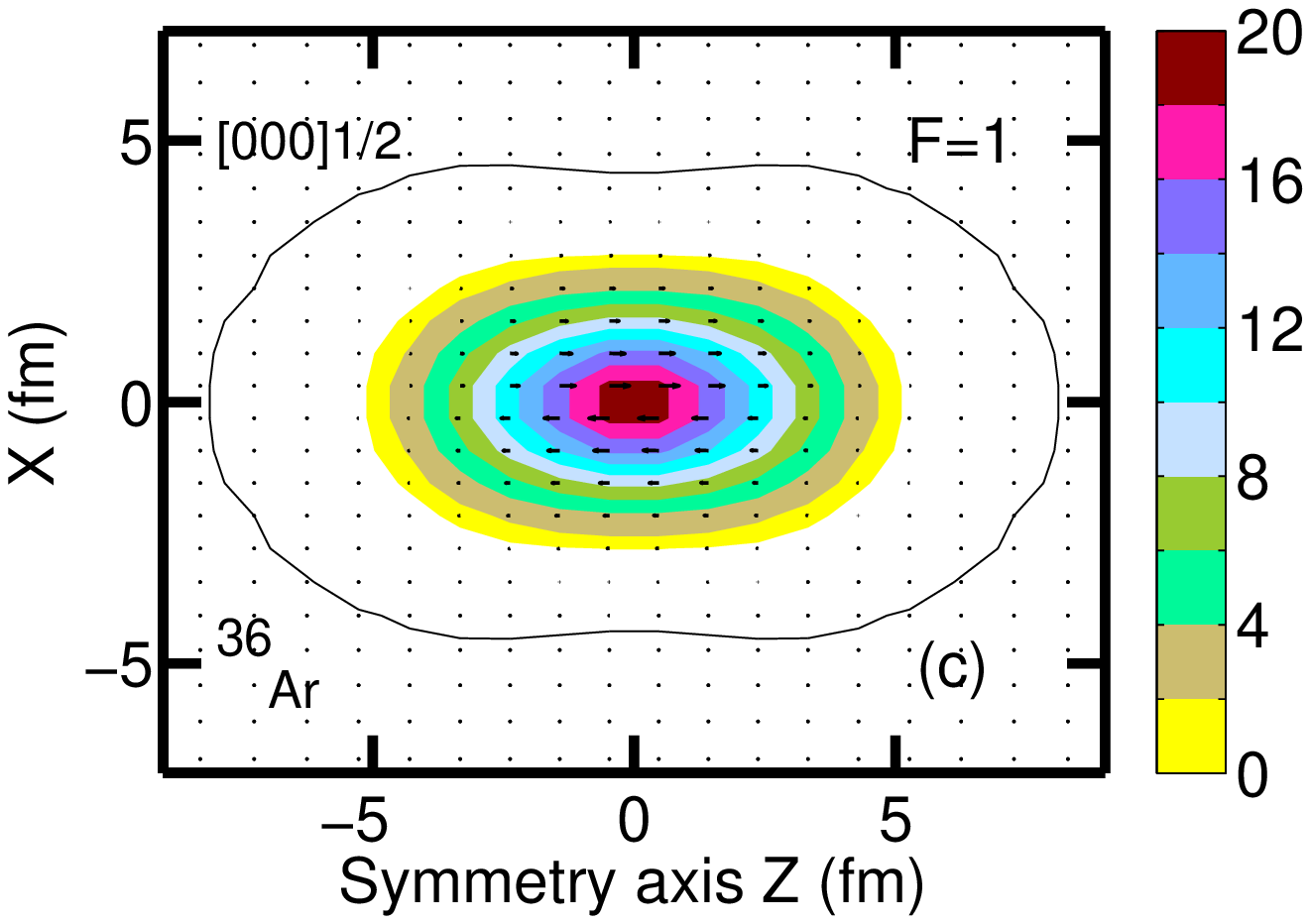} \\
\includegraphics[angle=0,width=5.45cm]{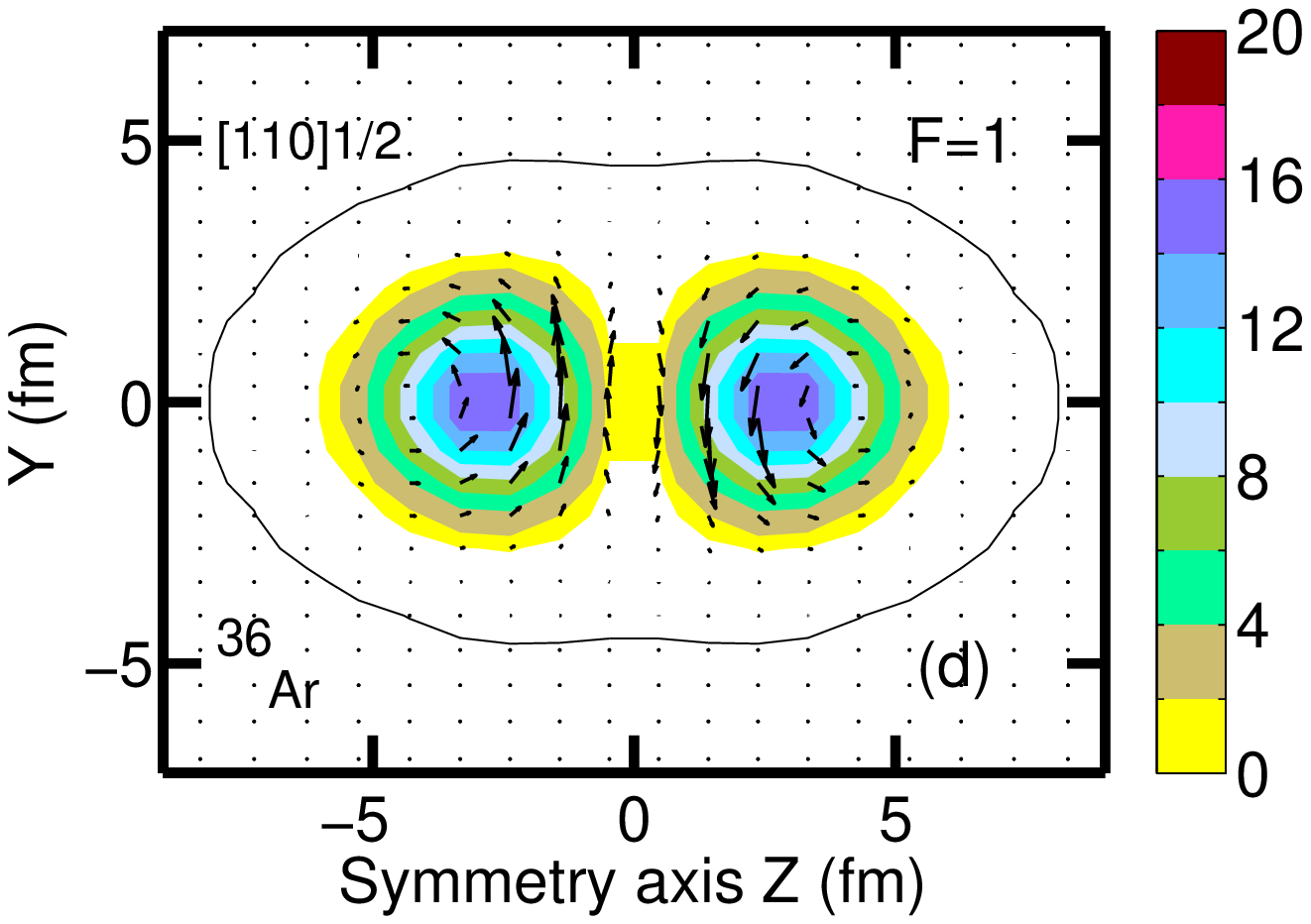}
\includegraphics[angle=0,width=5.45cm]{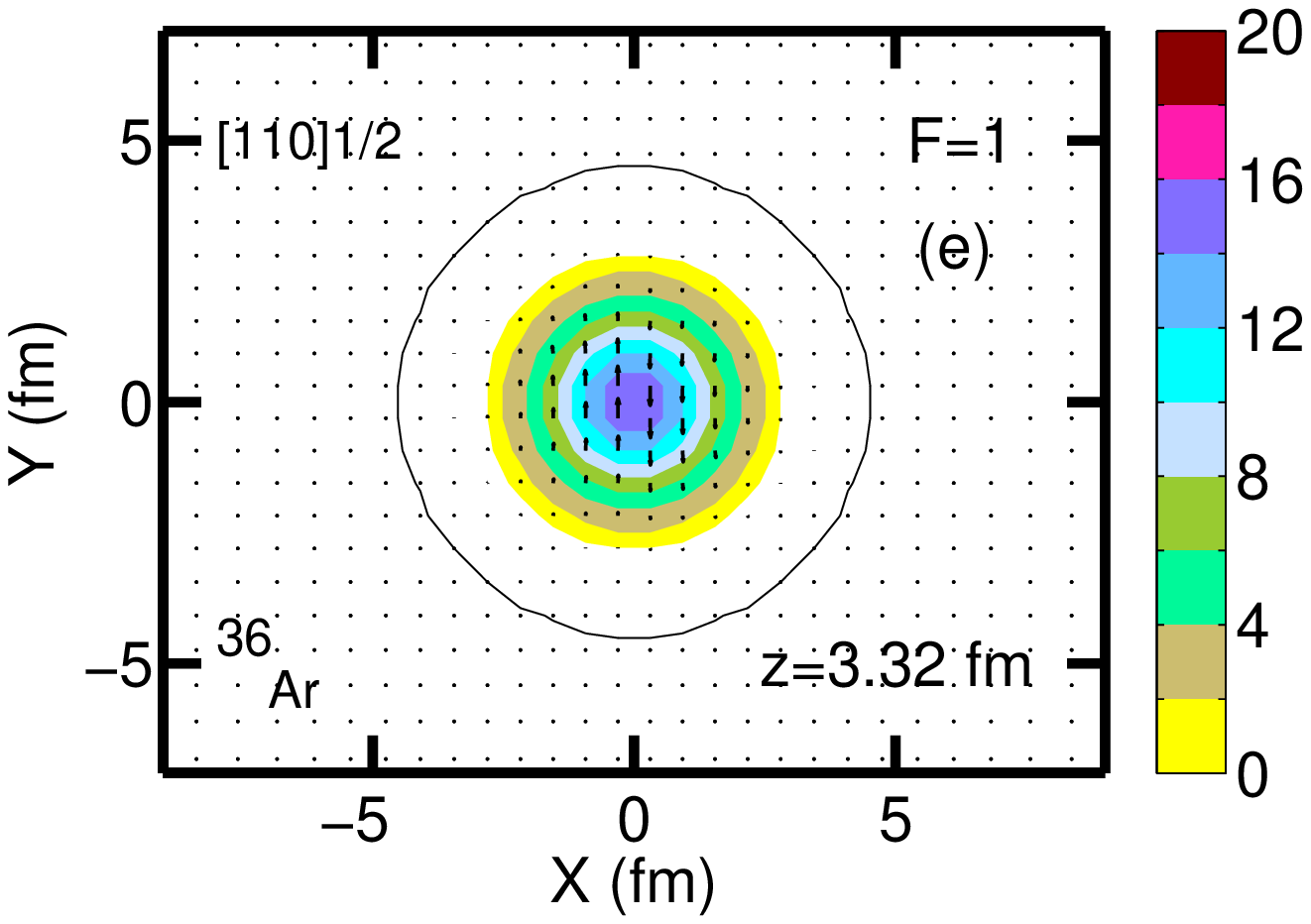}
\includegraphics[angle=0,width=5.45cm]{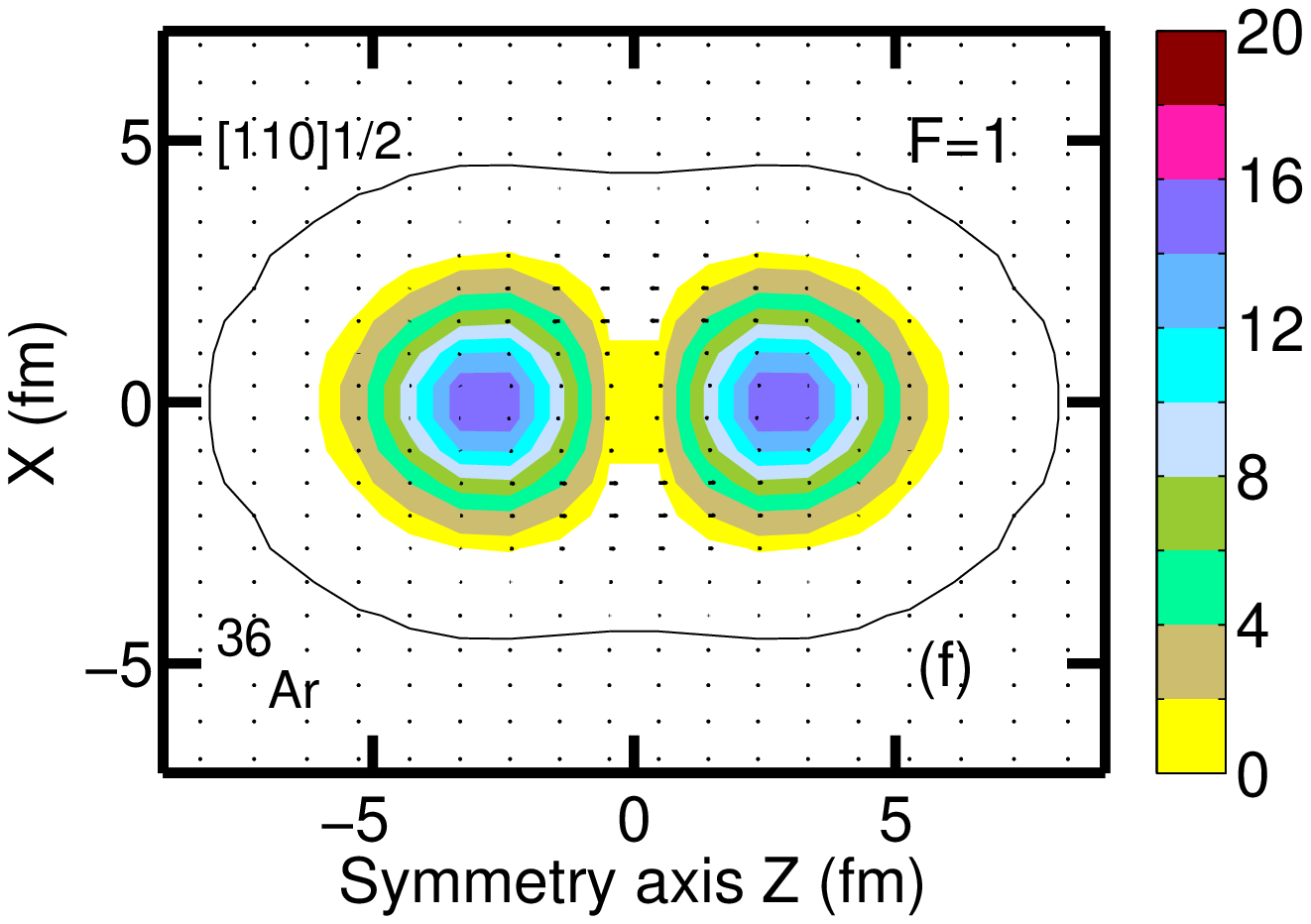} \\
\includegraphics[angle=0,width=5.45cm]{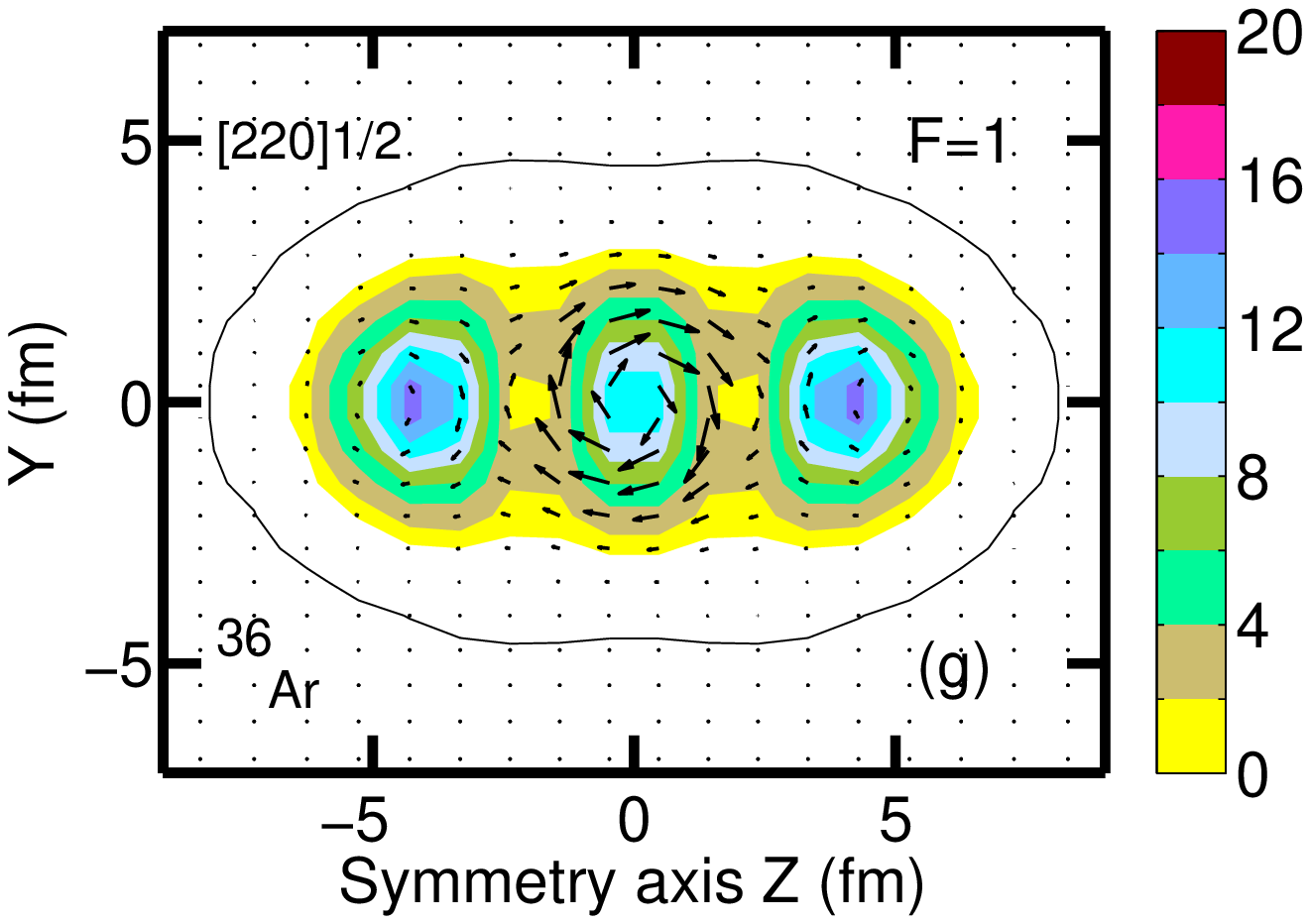}
\includegraphics[angle=0,width=5.45cm]{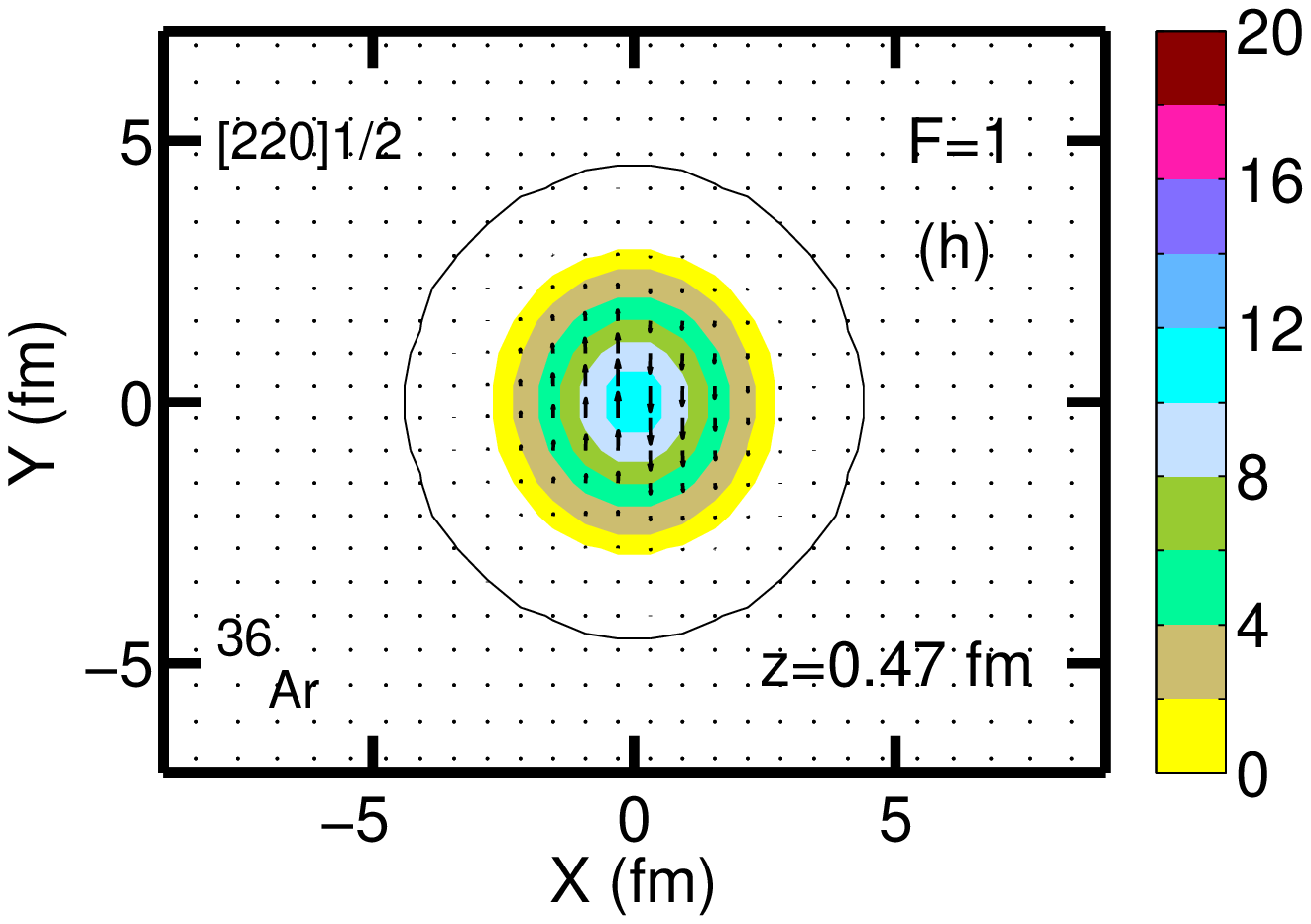}
\includegraphics[angle=0,width=5.45cm]{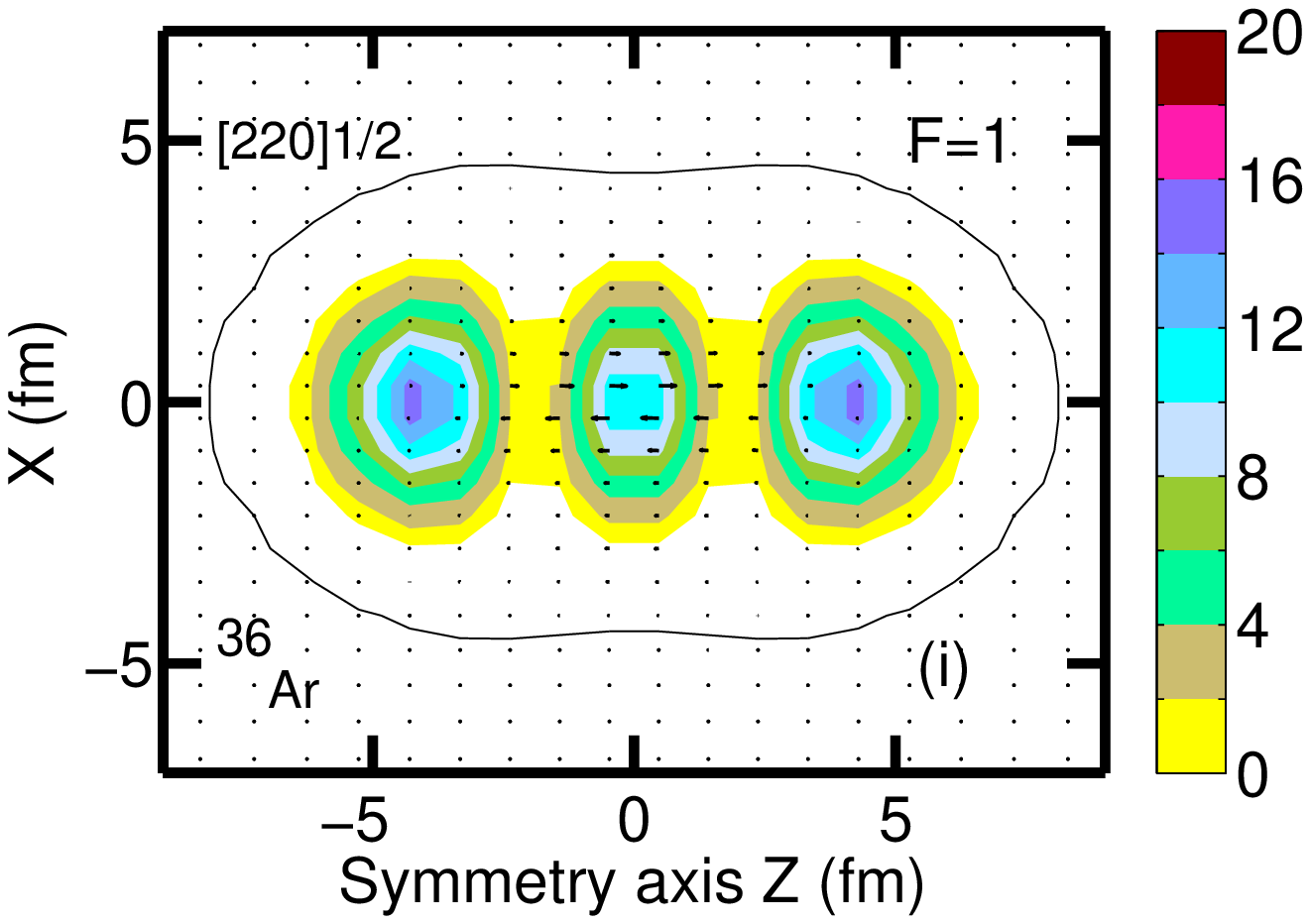} \\
\includegraphics[angle=0,width=5.45cm]{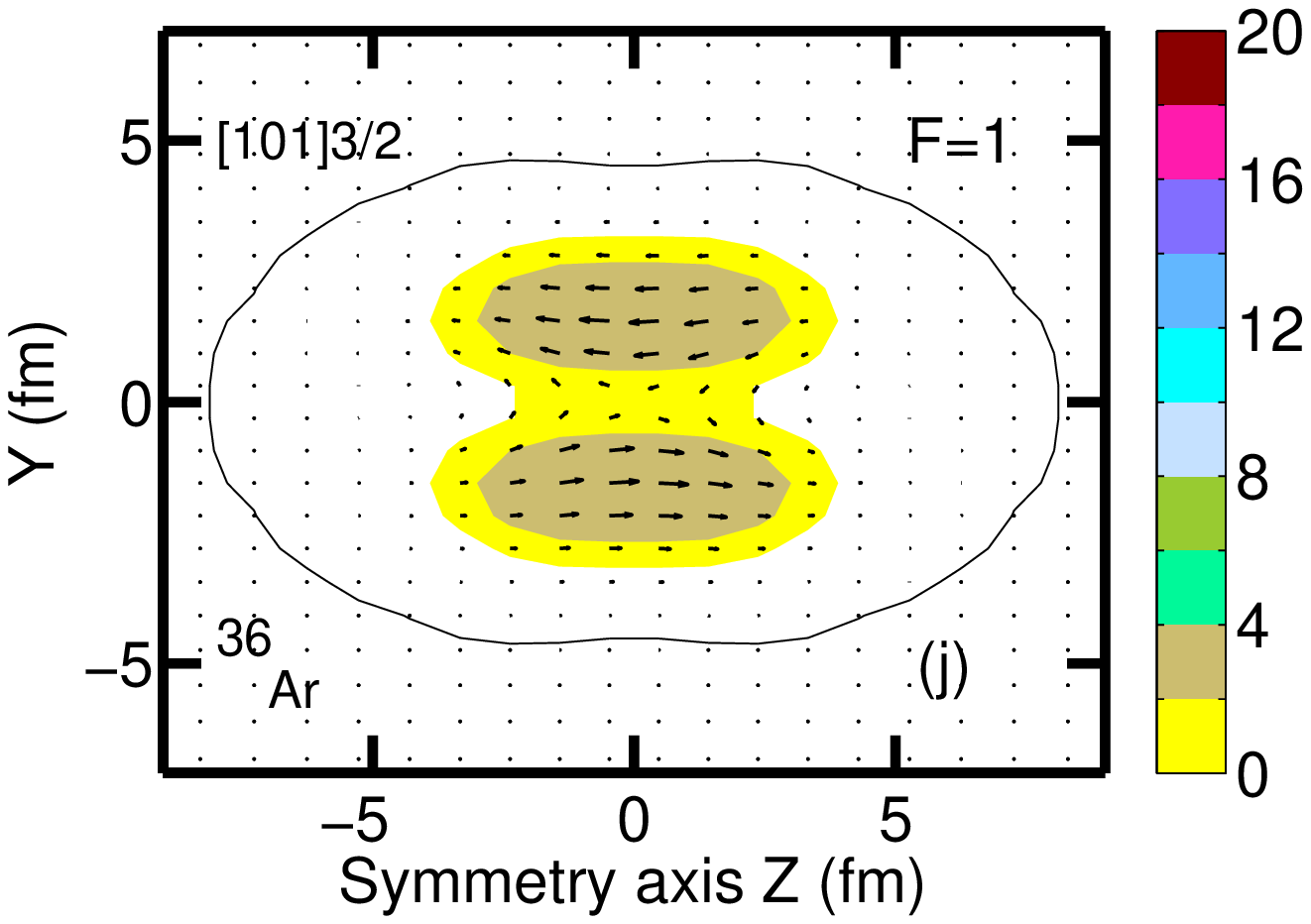}
\includegraphics[angle=0,width=5.45cm]{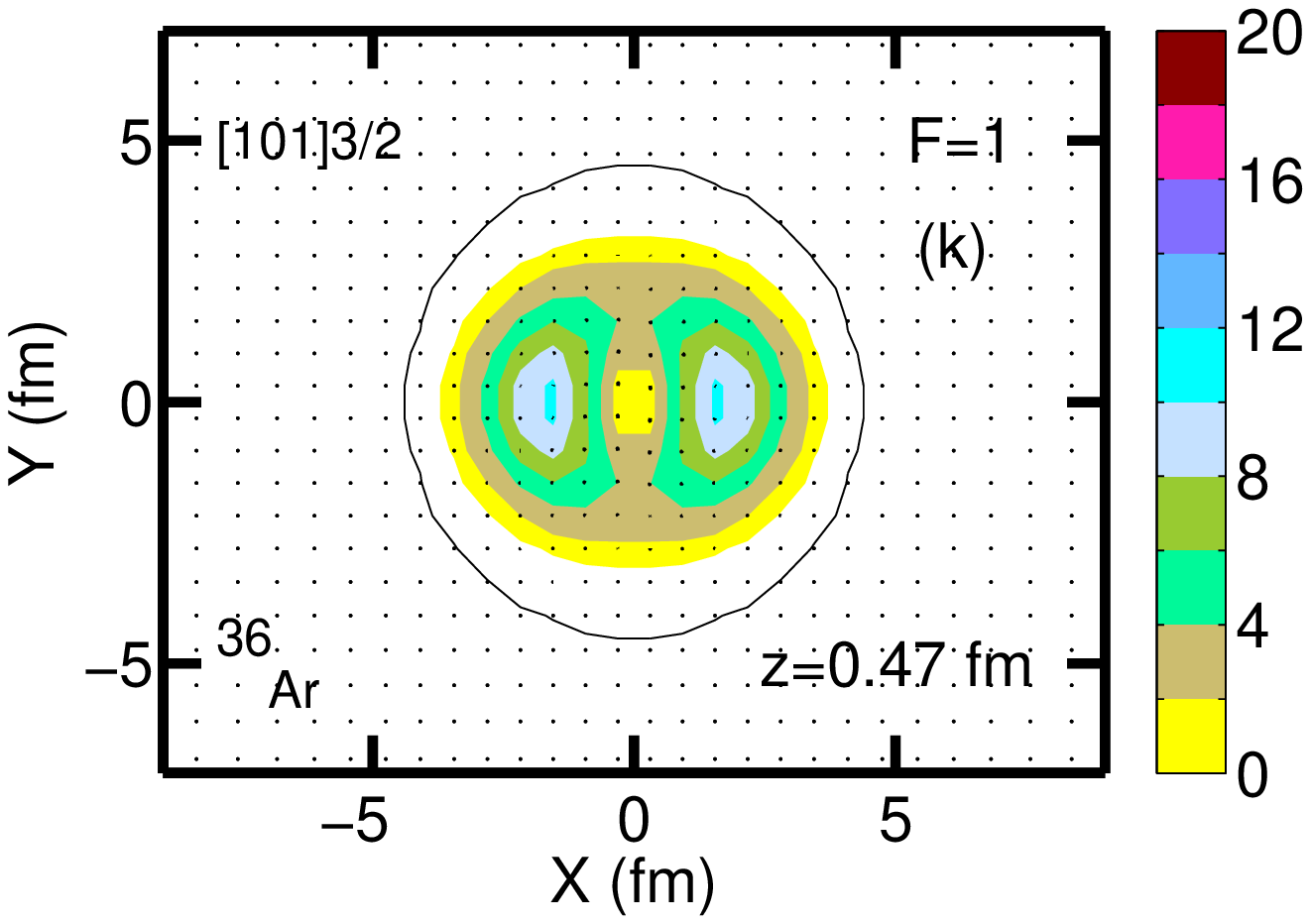}
\includegraphics[angle=0,width=5.45cm]{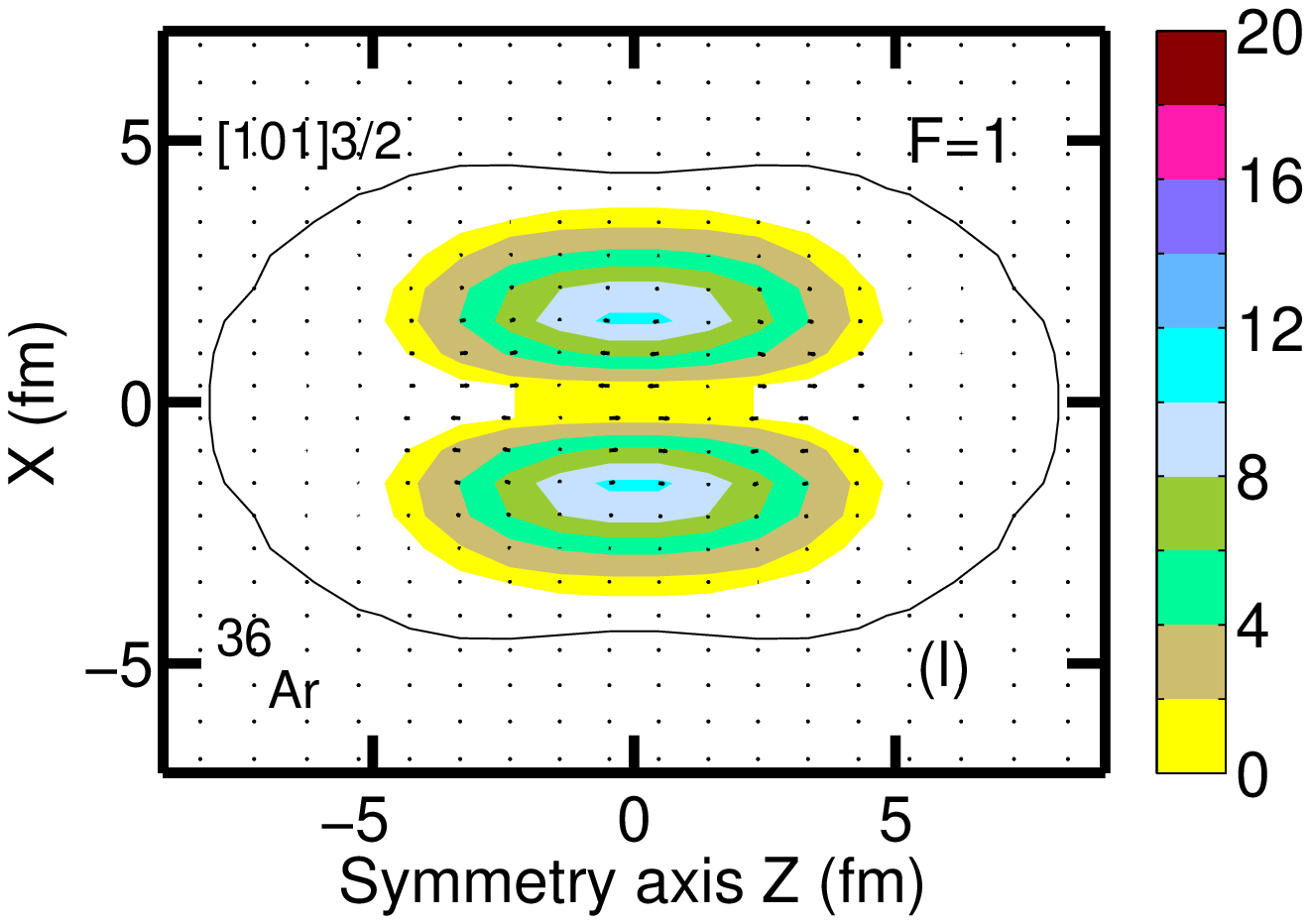} \\
\includegraphics[angle=0,width=5.45cm]{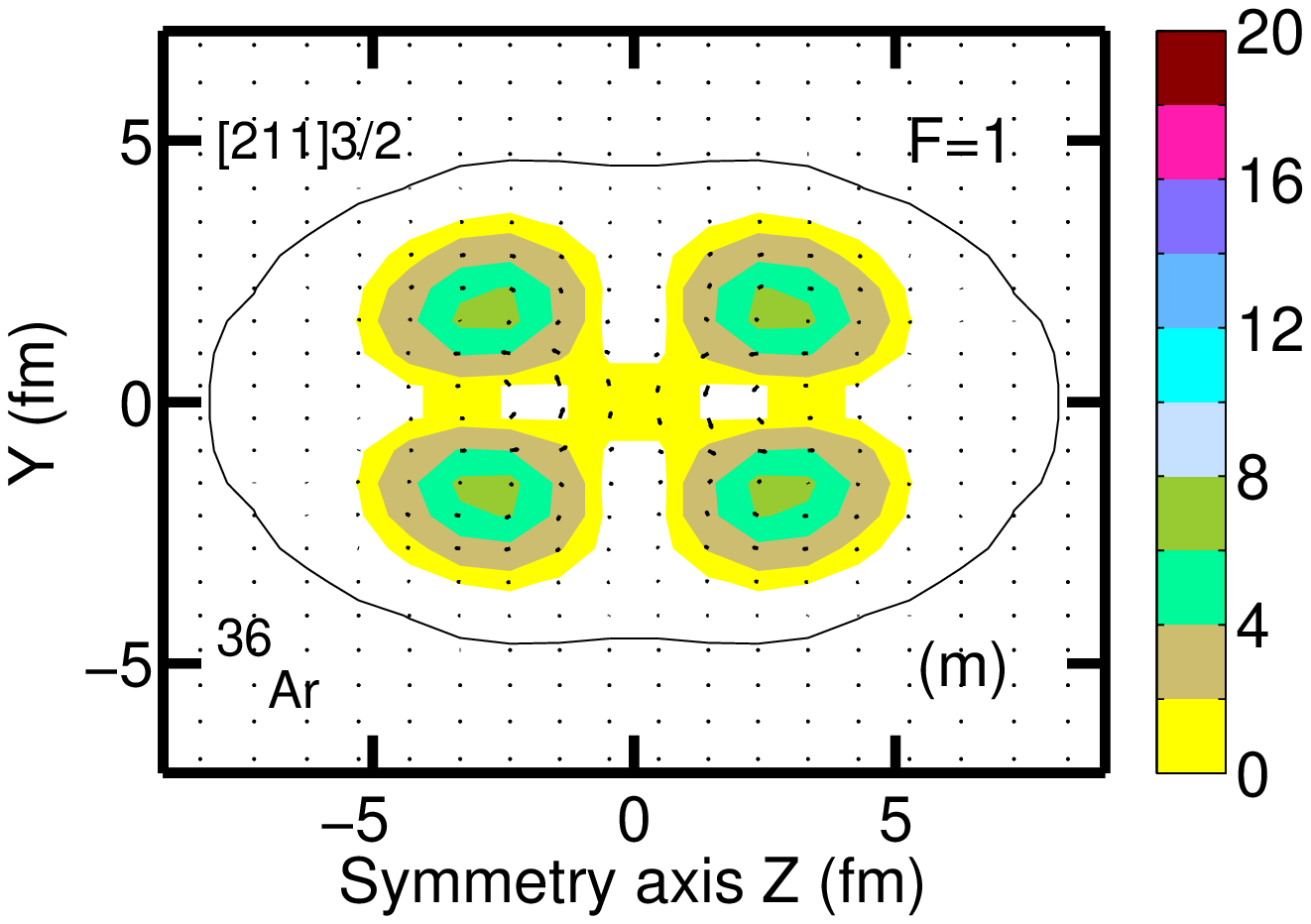}
\includegraphics[angle=0,width=5.45cm]{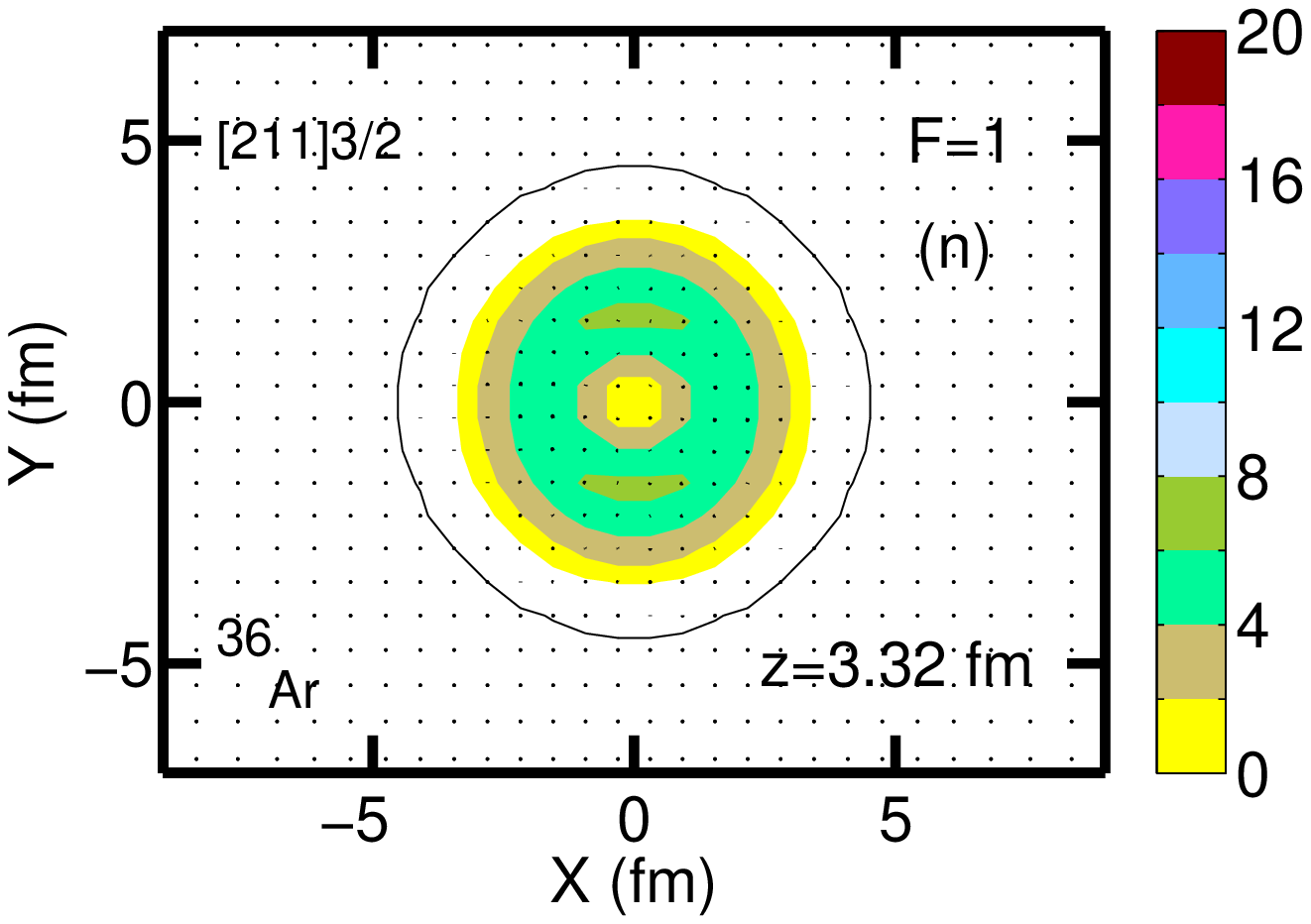}
\includegraphics[angle=0,width=5.45cm]{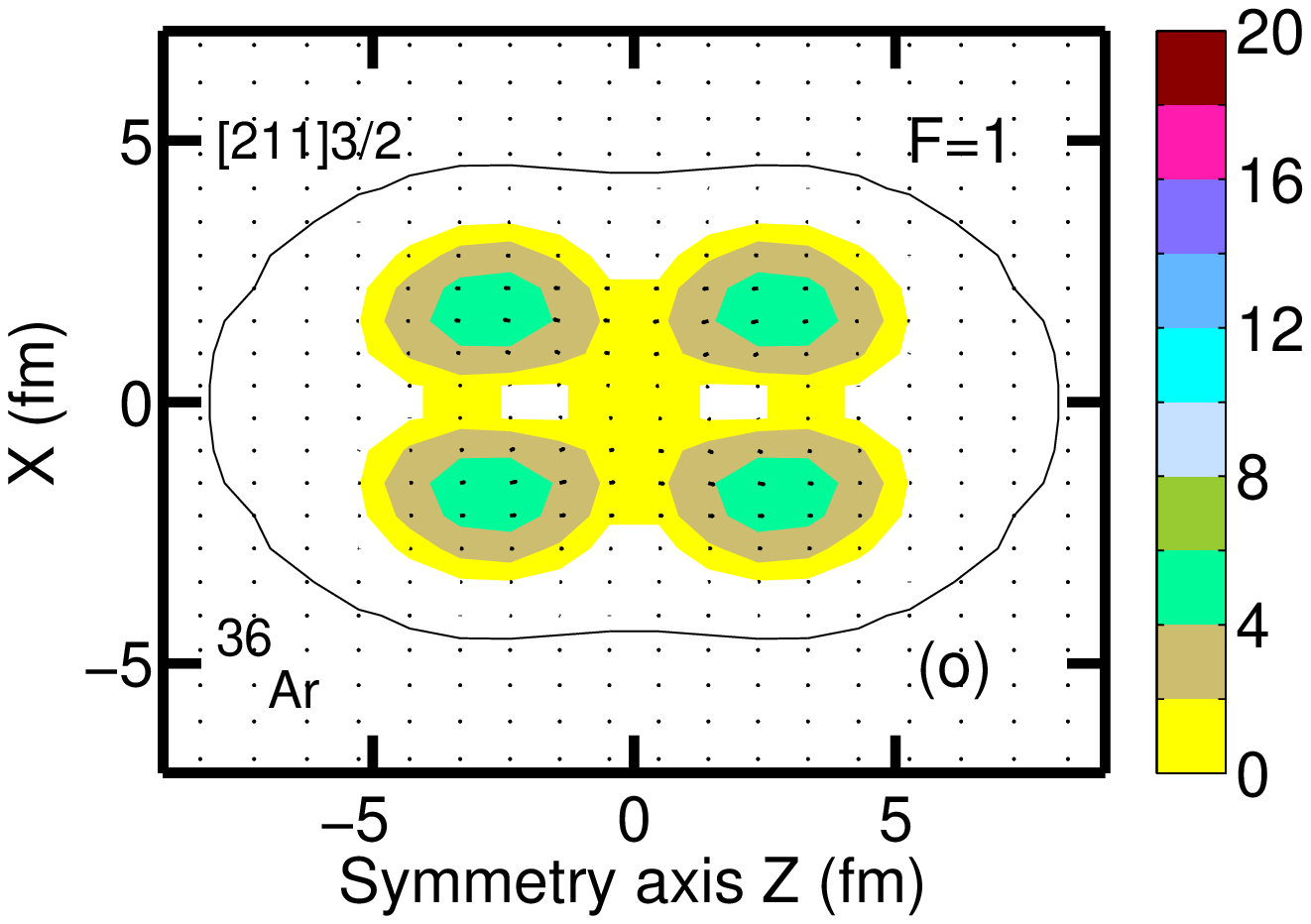}
\caption{The single-neutron density distributions [in 0.001 fm$^{-1}$] due to 
the occupation of the indicated Nilsson states with signature $r=-i$ in the 
megadeformed [31,31] configuration of $^{36}$Ar obtained in the calculations
with the NL3* CEDF. To give a full 3-dimensional representation of the 
single-particle density distributions, they are plotted 
in the $xz$ and $yz$ planes at the positions of the Gauss-Hermite integration 
points in the $y$ and $x$ directions closest to zero, namely, at $x=y=0.310$ fm, 
and in the $xy$ plane at the Gauss-Hermite integration point in the $z$-coordinate 
(the value of this coordinate is shown in middle panels) which gives the largest 
density. The states are shown from the bottom of nucleonic potential in the same 
sequence  as they appear in the routhian diagram of this configuration. The colormap 
shows the densities as multiplies of $0.001$ fm$^{-3}$. The shape and size of the nucleus 
are indicated by black solid line which corresponds to total neutron density of 
$\rho=0.001$ fm$^{-3}$. In addition, the current distributions {\bf j}$^n$({\bf r}) 
produced by these states are shown by arrows. The currents in panel (a) are plotted 
at arbitrary units for better visualization. In other panels they are normalized to 
the currents  in above mentioned panel by using factor F.
}
\label{Ar36-sp-p1}
\end{figure*}

\begin{figure*}[htb]
\includegraphics[angle=0,width=5.75cm]{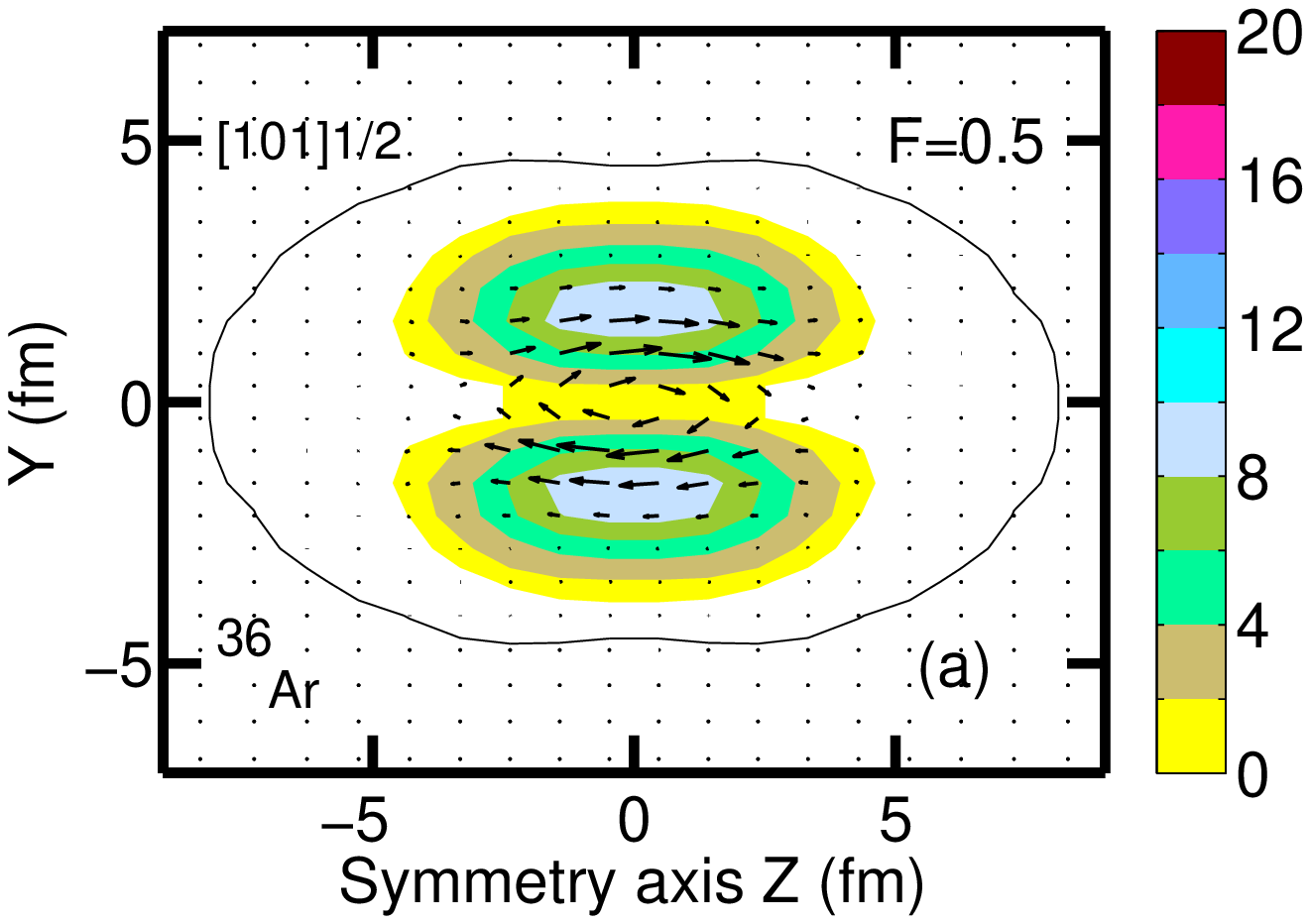}
\includegraphics[angle=0,width=5.75cm]{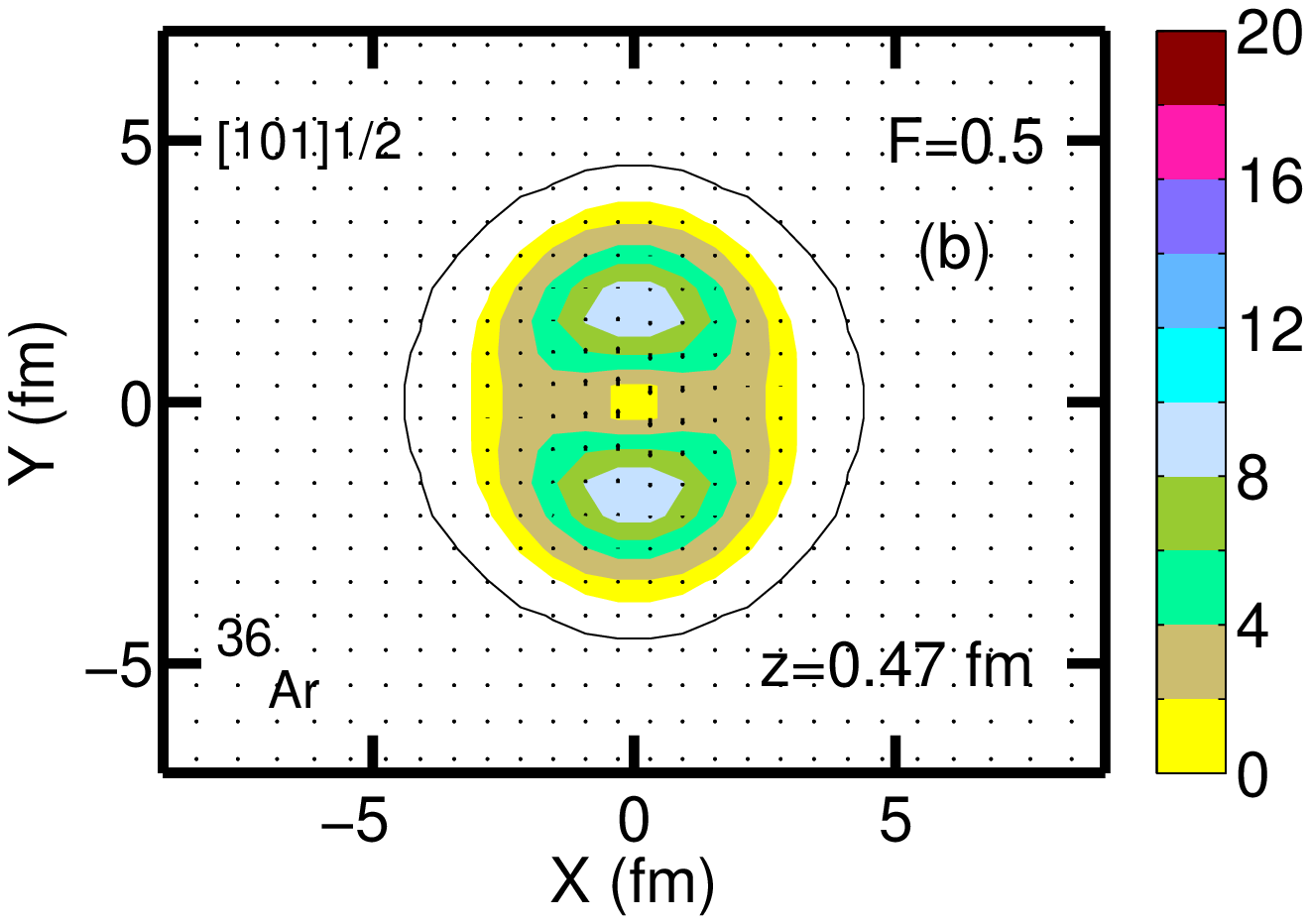}
\includegraphics[angle=0,width=5.75cm]{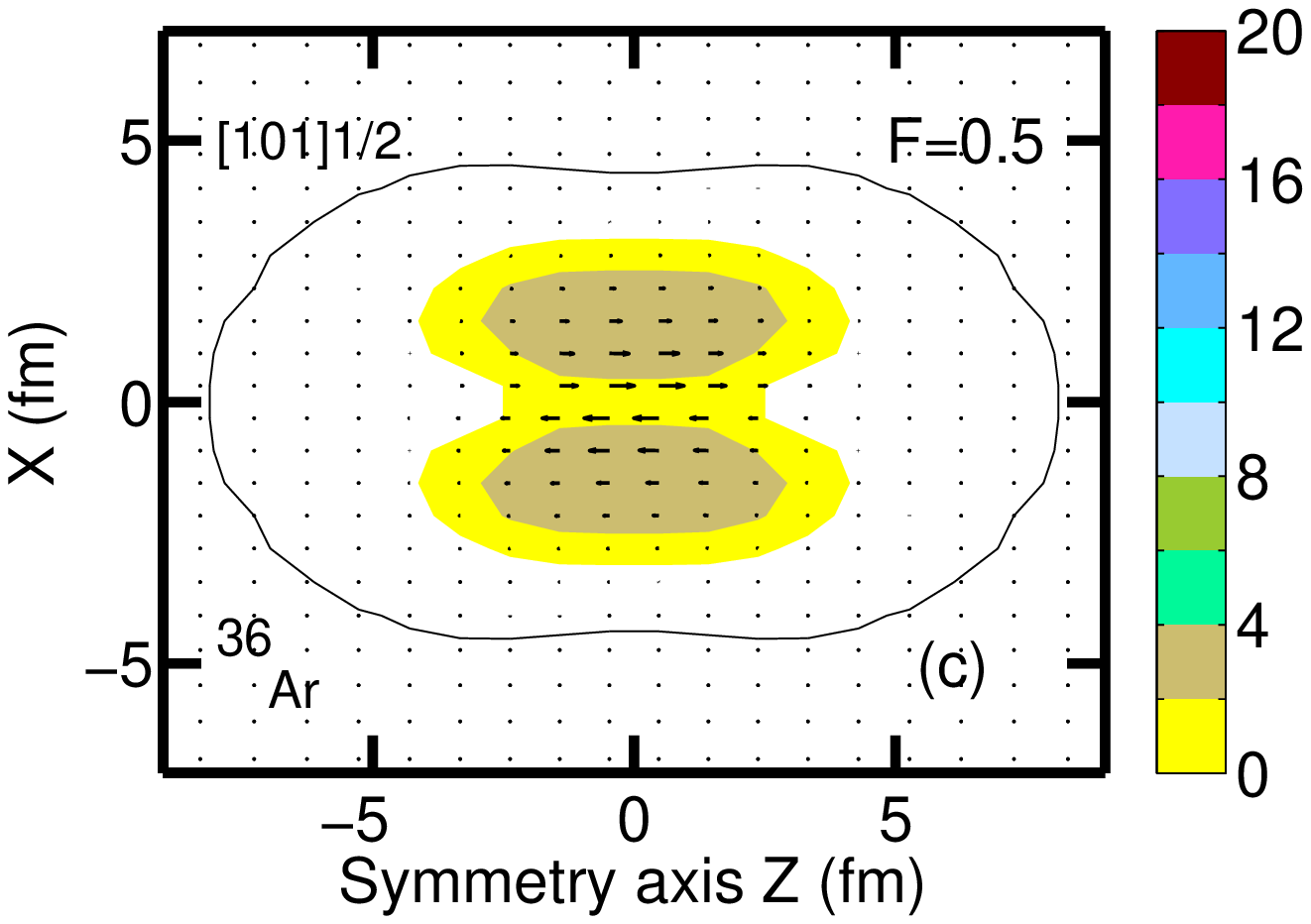} \\
\includegraphics[angle=0,width=5.75cm]{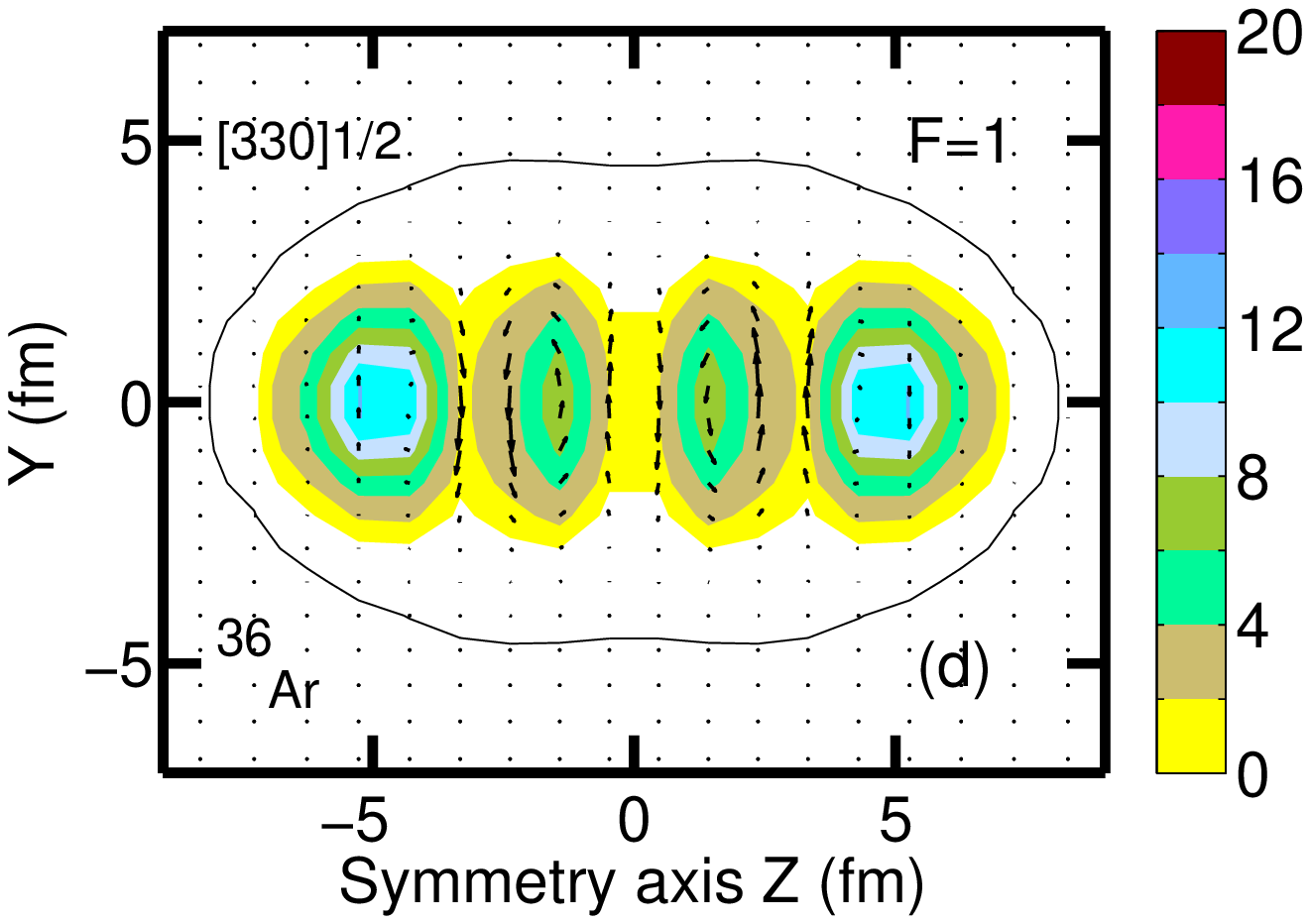}
\includegraphics[angle=0,width=5.75cm]{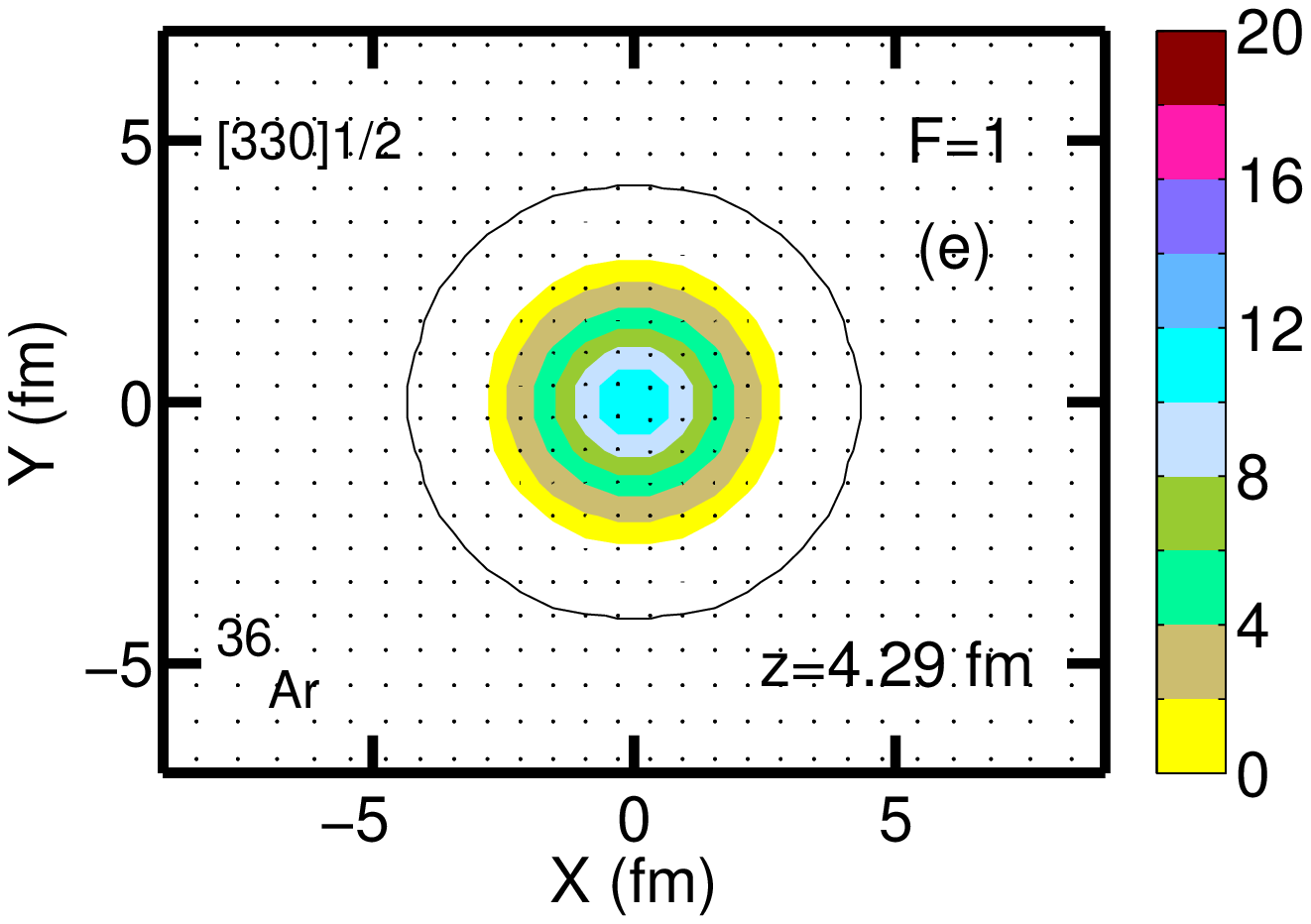}
\includegraphics[angle=0,width=5.75cm]{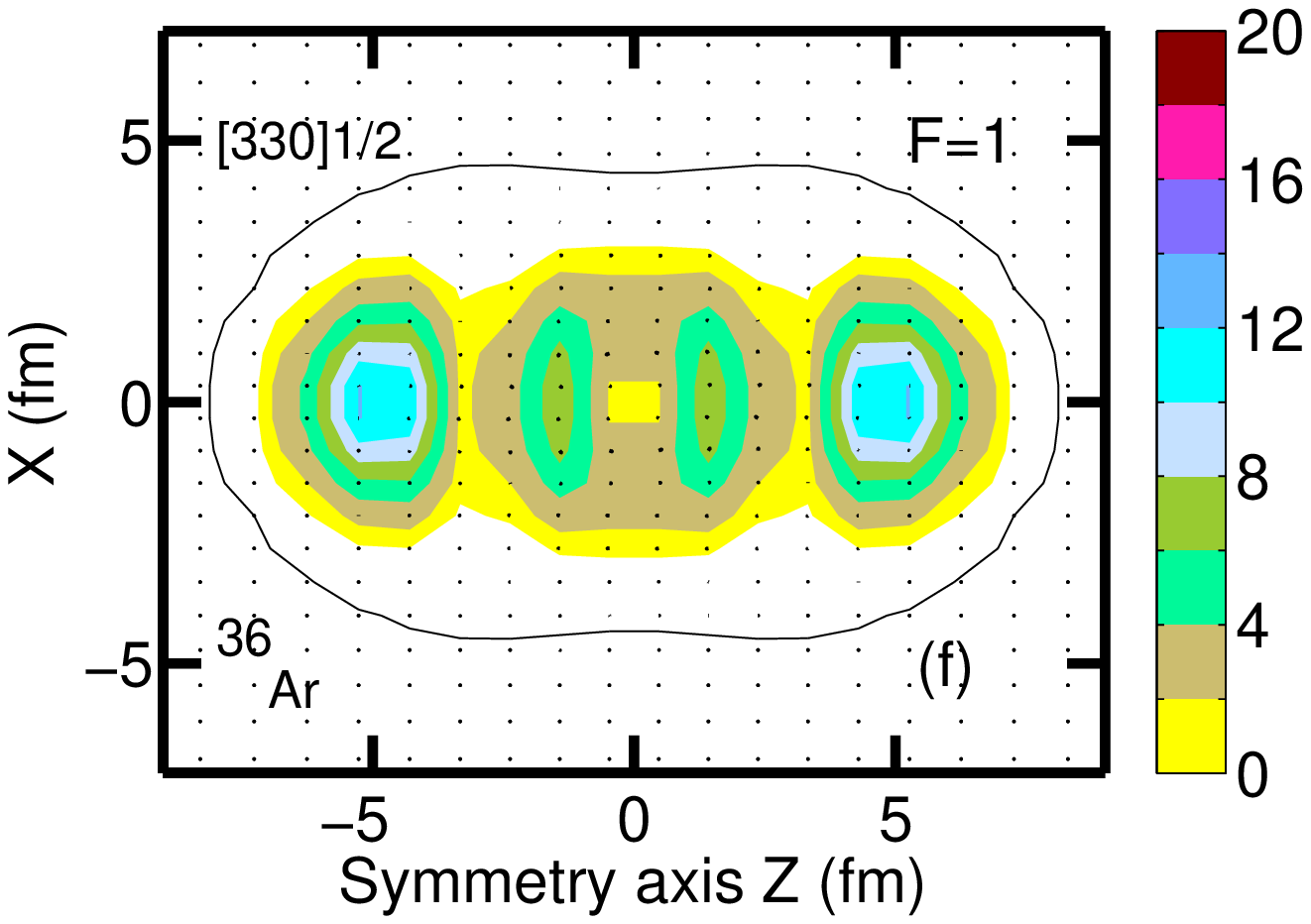} \\
\includegraphics[angle=0,width=5.75cm]{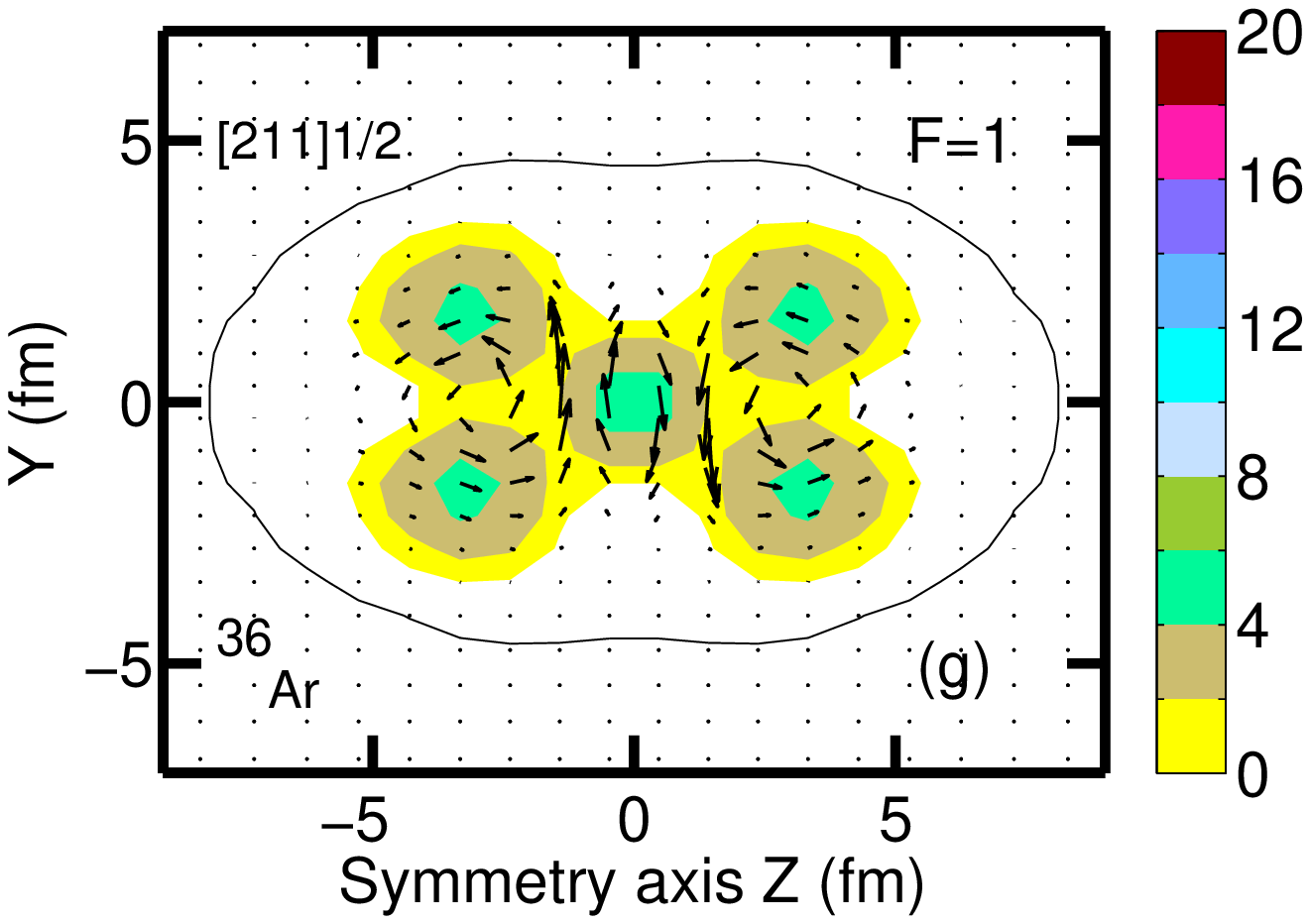}
\includegraphics[angle=0,width=5.75cm]{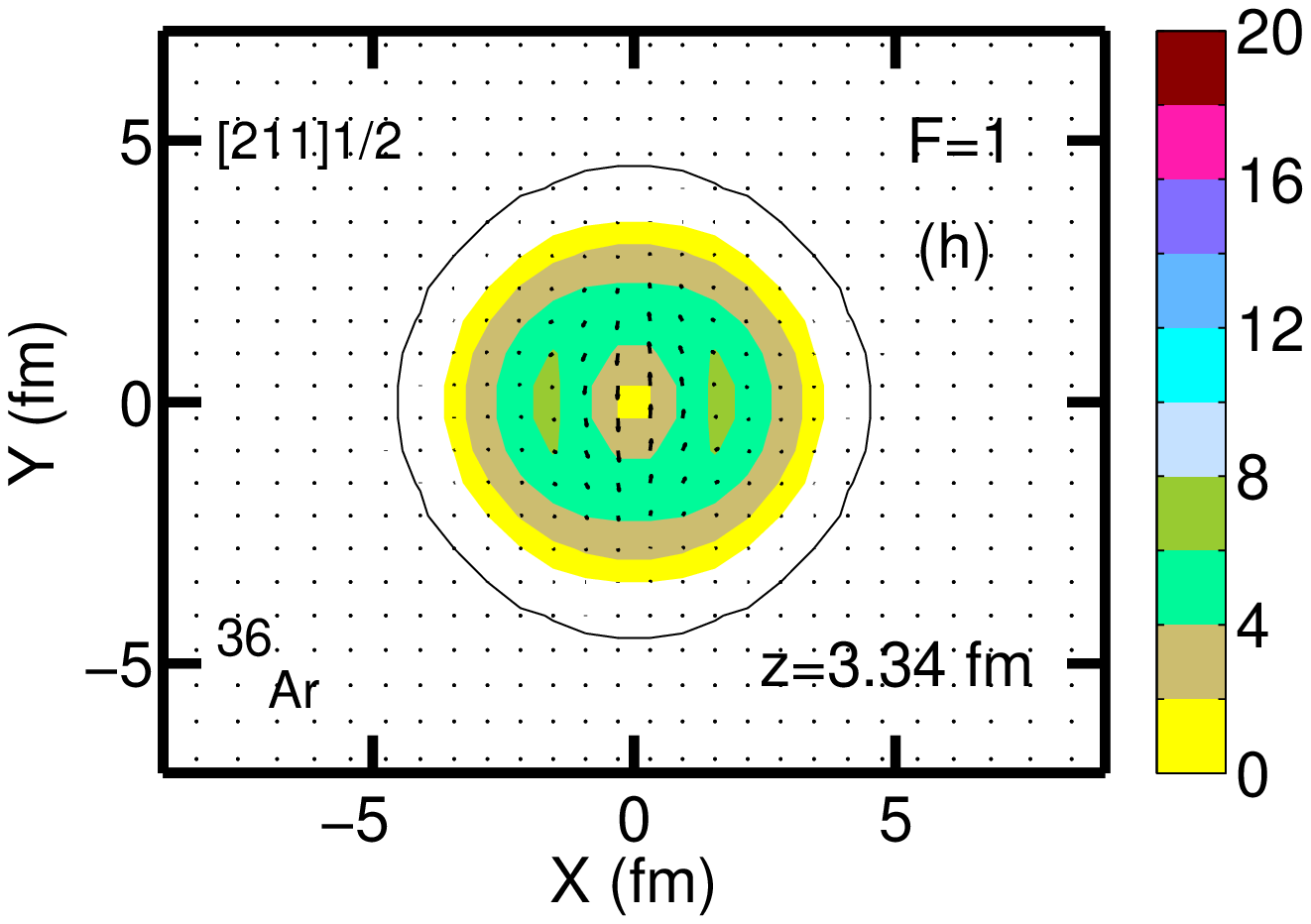}
\includegraphics[angle=0,width=5.75cm]{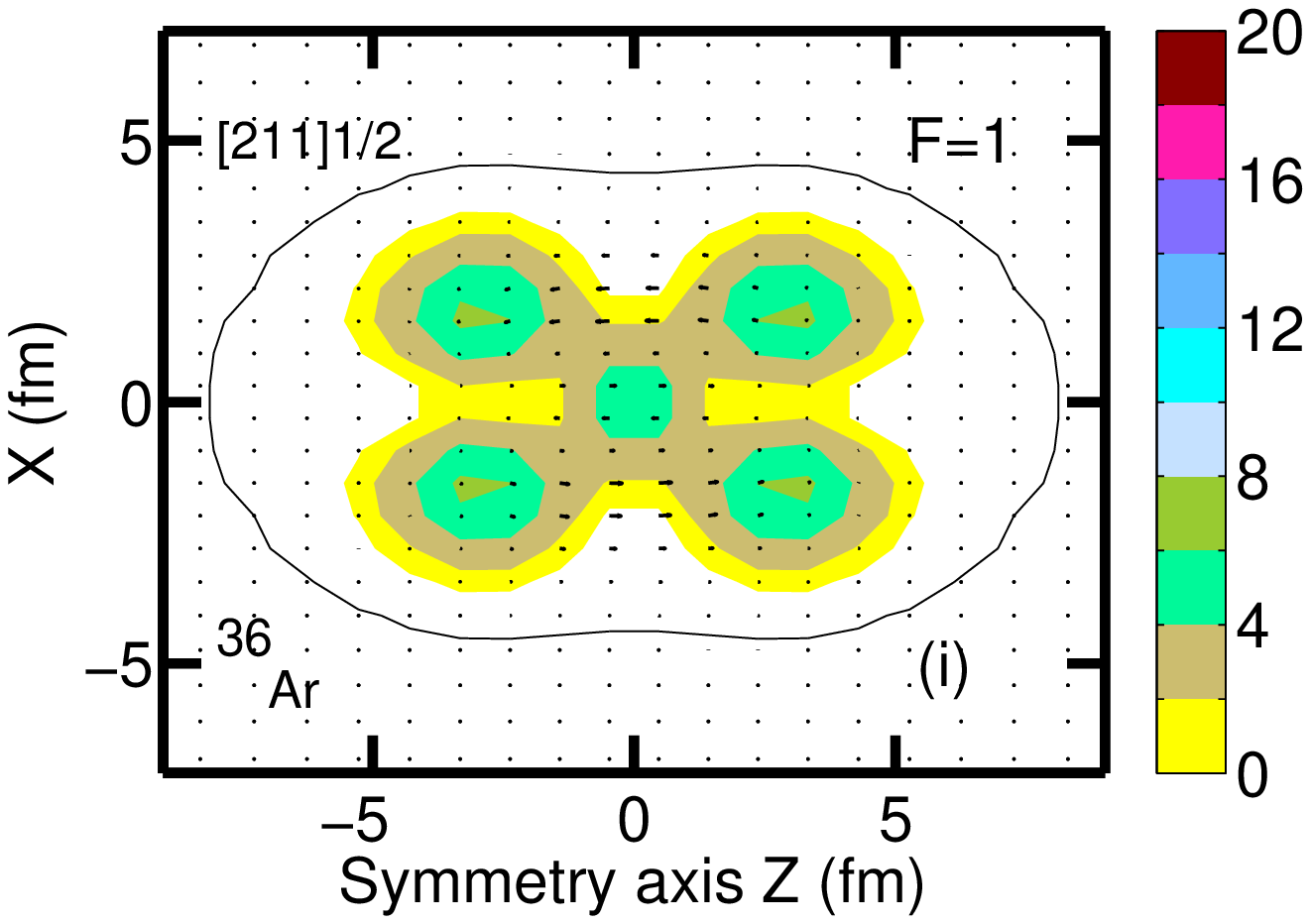} \\
\includegraphics[angle=0,width=5.75cm]{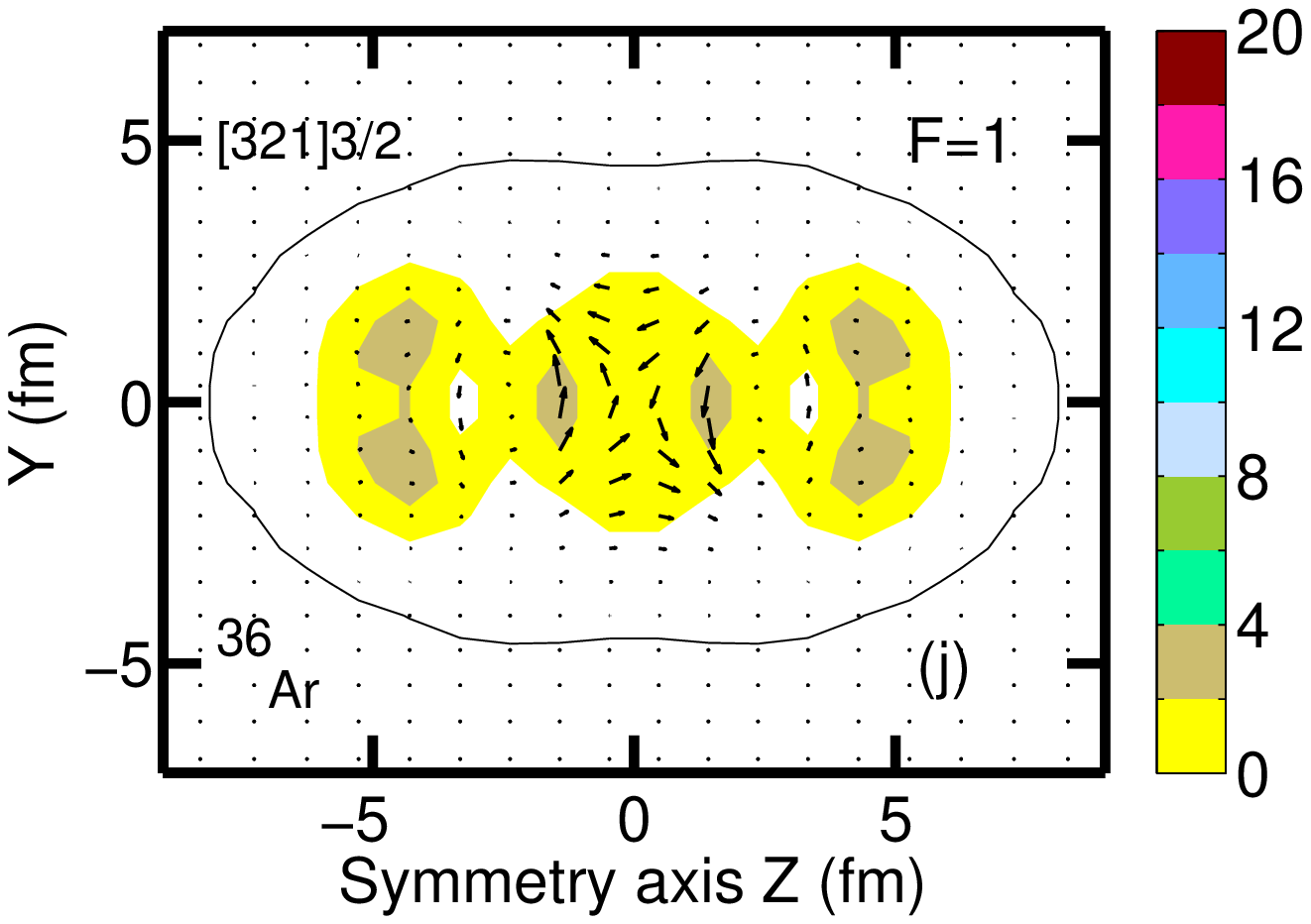}
\includegraphics[angle=0,width=5.75cm]{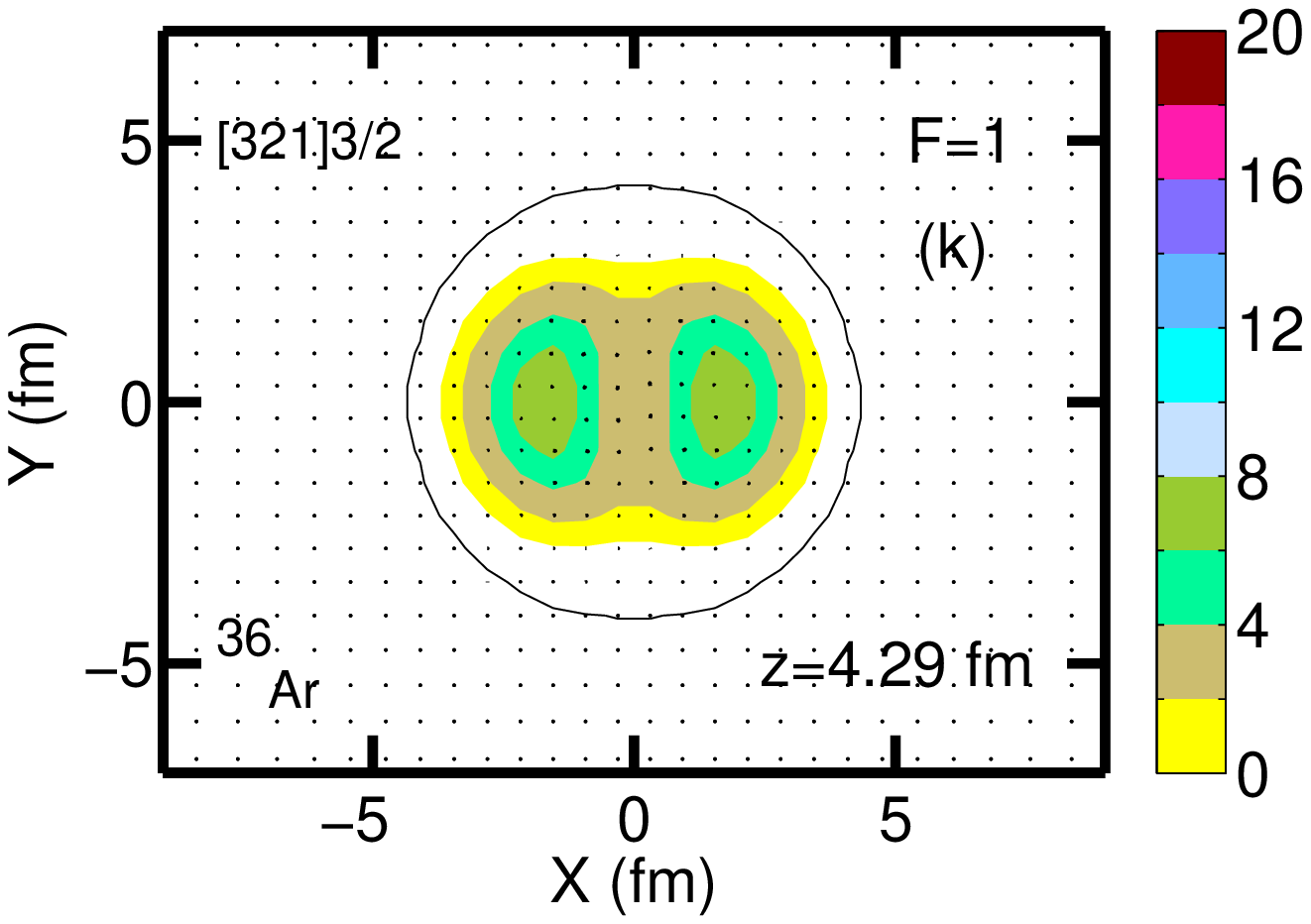}
\includegraphics[angle=0,width=5.75cm]{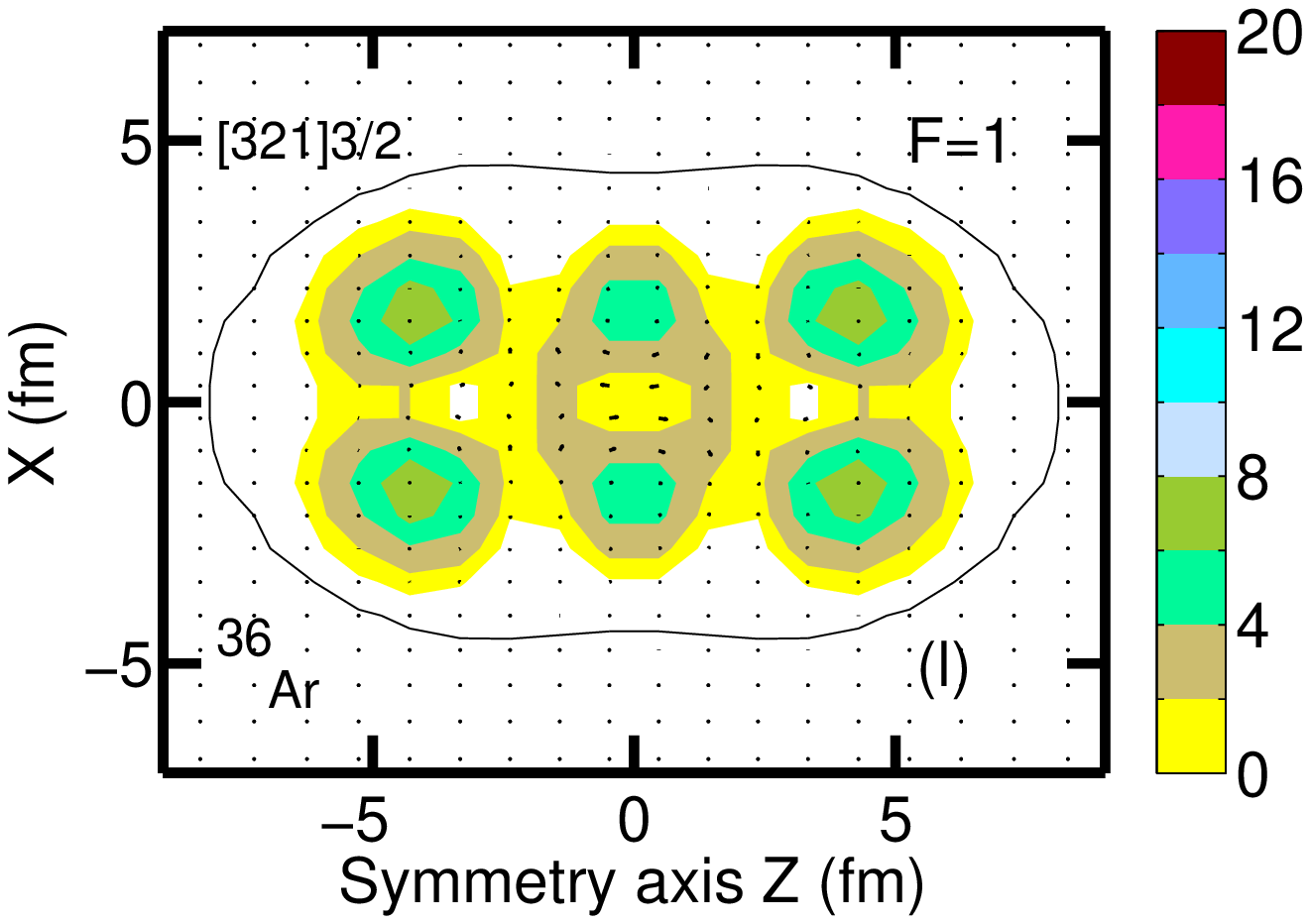} \\
\includegraphics[angle=0,width=5.75cm]{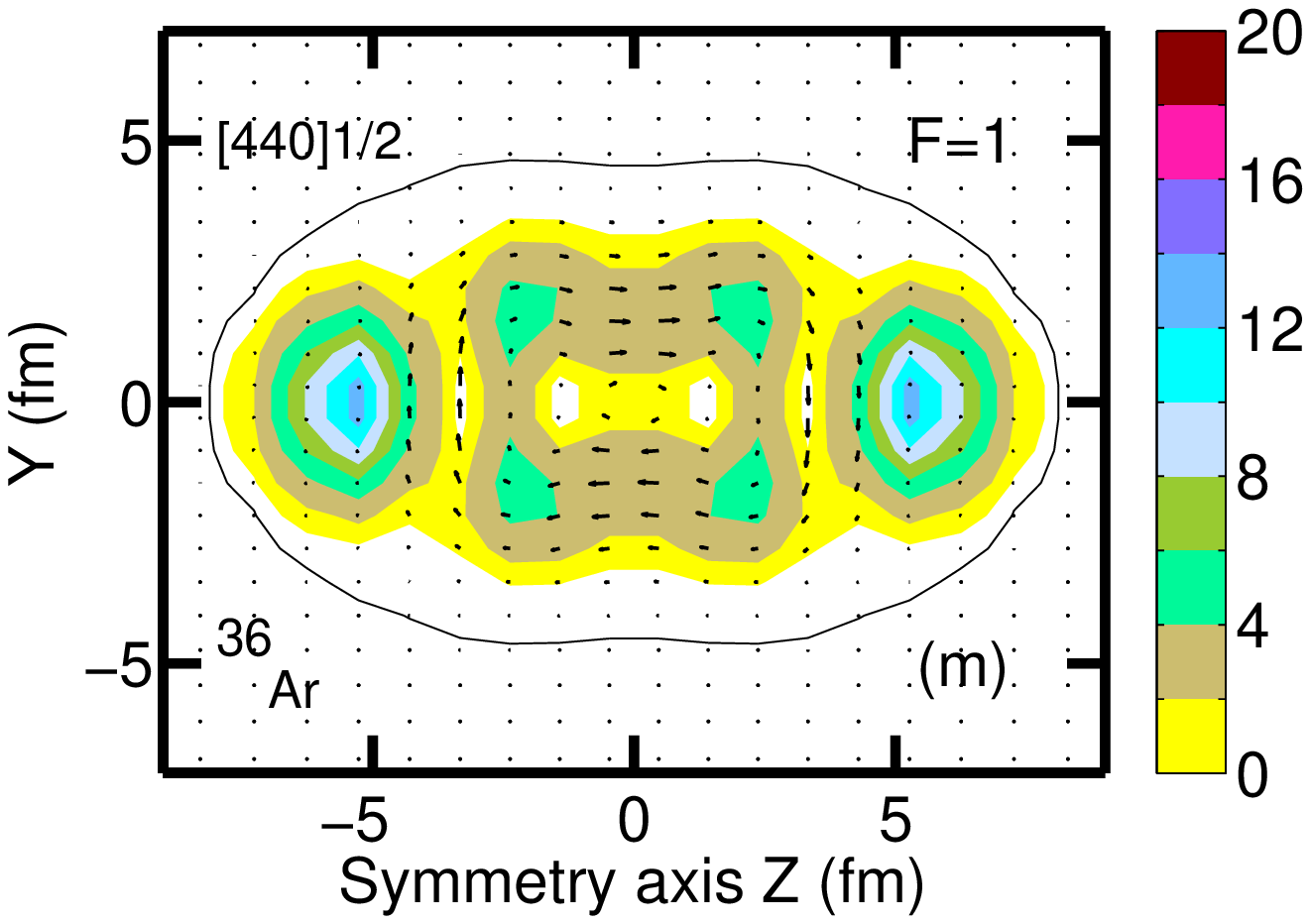}
\includegraphics[angle=0,width=5.75cm]{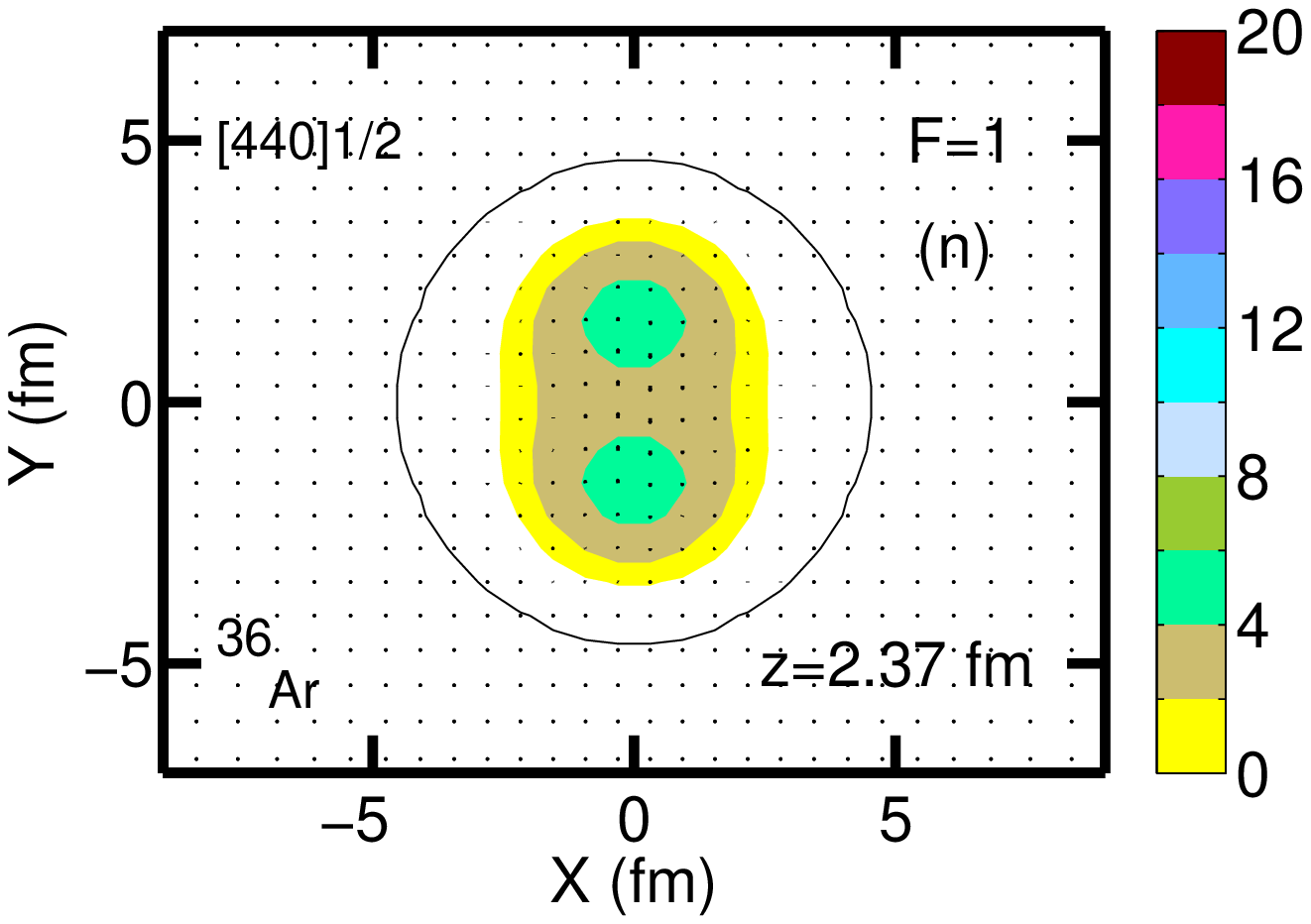}
\includegraphics[angle=0,width=5.75cm]{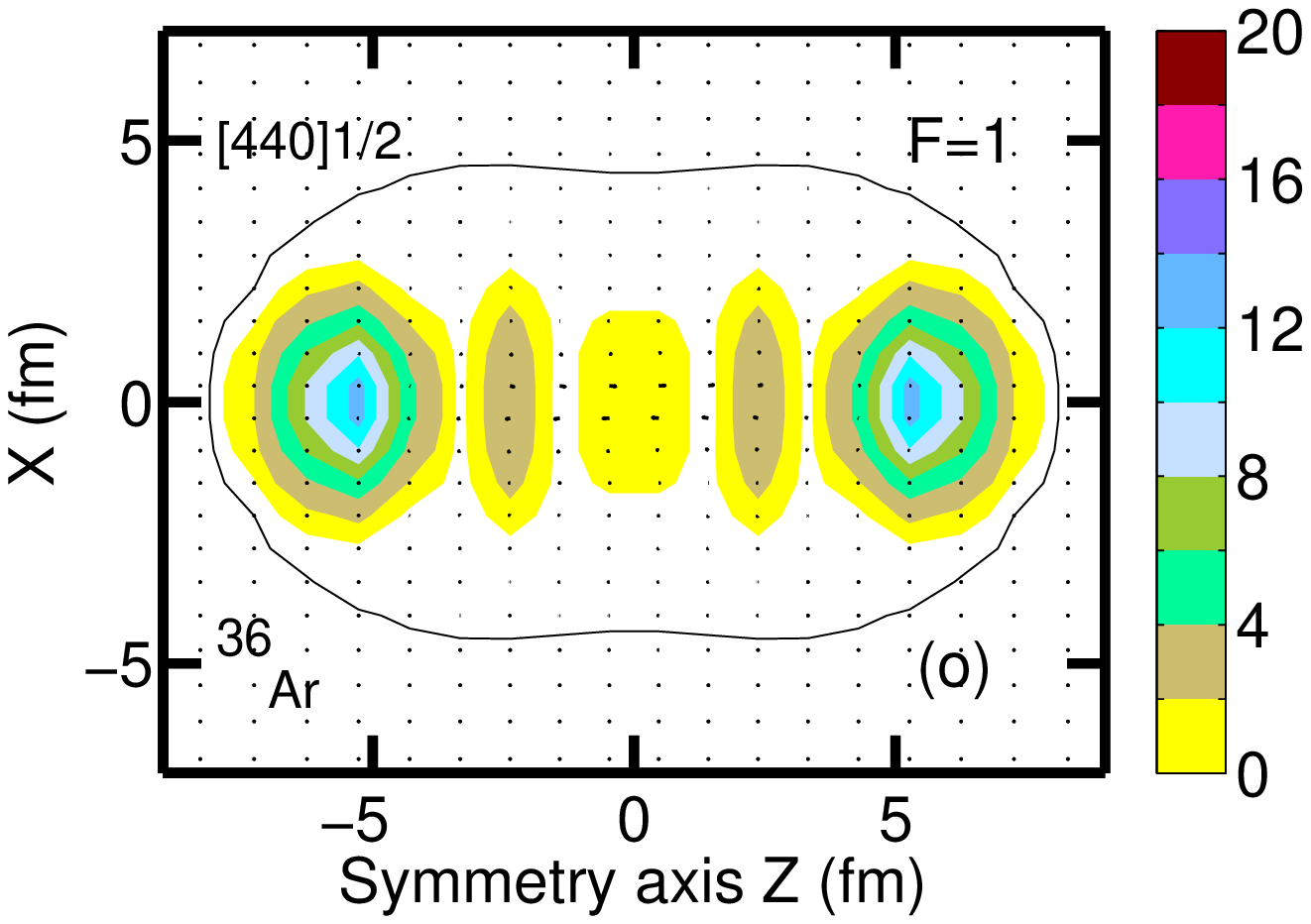}
\caption{The same as Fig.\ \ref{Ar36-sp-p1} but for remaining occupied 
neutron orbitals with $r=-i$ in the megadeformed [31,31] configuration 
of $^{36}$Ar.}
\label{Ar36-sp-p2}
\end{figure*}

  The calculations in the cranked relativistic mean field (CRMF) \cite{RA.16} 
and cranked Nilsson-Strutinsky (CNS) \cite{Pingst-A30-60} frameworks clearly 
indicate $^{36}$Ar as one of the best candidates for the observation of the hyper- 
(HD) and megadeformation (MD) at high spin. The observed superdeformed band terminates 
at spin $I=16^+$ \cite{36Ar}. The population of the HD and MD states is very likely 
if it will be possible to bring higher (than $16\hbar$) angular momentum into the 
system \cite{RA.16}. For example, the MD [31,31] configuration is predicted to 
become yrast at spin $I\geq 21\hbar$ (see Fig. 21 in Ref.\ \cite{RA.16}). Here the 
calculated configurations are labeled by shorthand [$n_{1}$$n_{2}$, $p_{1}$$p_{2}$] 
labels, where $n_{1}$ and $n_{2}$  ($p_{1}$ and $p_{2}$) are the number of neutrons 
(protons) in the $N=3$ and 4 intruder/hyperintruder orbitals.

  Fig.\ \ref{Total-densities} shows the total neutron density distribution of 
this configuration at the spin at which this configuration becomes yrast.  The
proton density distribution is almost the same; thus it is not shown here. One 
can see clear fingerprints of the molecular structure in this density distribution; 
two clusters with high densities in their near-central region are separated by a 
well-established neck. It looks as a pair of two octupole (pear-shaped) deformed 
$^{18}$F nuclei. This is one of the forms of the clusterization predicted in 
nuclei \cite{OFE.06}. It is reasonable to expect that single-particle degrees of 
freedom play an important role in the formation of molecular structures. However, 
to our knowledge this question has never been studied in detail in nuclei with 
$A\geq 20$ and in rotating nuclei. This is contrary to the situation in very light 
nuclei in which the connection between $\alpha$ clusterization and underlying 
single-particle structure has been explored for non-rotating nuclei in detail 
in a number of publications (see, for example, Refs.\ \cite{AJ.94,FBW.95,EKNV.14,YIM.14}). 
For example, the buildup of total nucleonic density of the $\alpha$ cluster structures 
in the Be and C isotopes by means of the single-particle contributions has been 
explored within the CDFT framework in Ref.\ \cite{EKNV.14}.

  To better understand the role of the single-particle states and their nodal
structure in the buildup of total nucleonic density in molecular states we 
consider the density distributions of the neutron states with signature $r=-i$
occupied in the MD [31,31] configuration of $^{36}$Ar. The calculations are 
performed in the CRMF framework \cite{VALR.05} using the NL3* CEDF \cite{NL3*}
and their results are shown in Figs.\ \ref{Ar36-sp-p1} and \ref{Ar36-sp-p2}.
The one-dimensional rotation in the CRMF framework is along the $x$-axis \cite{VALR.05}. 
Note that the structure of the yrast and near-yrast states in $^{36}$Ar has been 
studied in detail in Ref.\ \cite{RA.16}. 
In addition, the current distributions {\bf j}$^n$({\bf r}) produced by 
these states are shown by arrows in Figs. \ref{Ar36-sp-p1} and 
\ref{Ar36-sp-p2}. As discussed in detail in Ref.\ \cite{TO-rot}), these 
currents have a significant impact on rotational properties of the nuclei.

 The single-particle orbitals are labeled by the asymptotic quantum numbers
$[Nn_z\Lambda]\Omega$ (Nilsson quantum numbers) of the dominant component of
the wave function. The shape of the [31,31] MD configuration is nearly axial 
with large quadrupole $\beta_2$ deformation (Fig. 23 in Ref.\ \cite{RA.16}). 
As a result, the weight of the dominant component exceed 75\% of the total 
wavefunction for the majority of the states. The only exceptions are the 
[440]1/2, [330]1/2 and [321]3/2 states for which the weights of the dominant 
component are 55\%, 62\% and 54\%, respectively.

  The single-particle states can be separated into several groups according 
to general features of their density distribution. One of the groups is 
represented by the $[NN0]1/2$ states for which the maximum of the density 
distribution in the density clusters is located at the axis of symmetry. 
The density clusters are spheroidal or elipsoidal in shape and the wavefunction 
does not have nodes in the direction perpendicular to the symmetry axis. The 
number of the density clusters in these states is equal to 
$N+1$ and the maximum density is always observed in the density clusters which 
are located in the polar region of the nucleus. Note that the maximum density in the 
clusters decreases with the increase of $N$. The wave function is well 
localized in such states with $N=0, 1$ and 2 [Figs.\ (\ref{Ar36-sp-p1})a-f]; 
among all considered single-particle states these are the ones with the 
highest densities in the center of the density clusters. This is a reason 
why they play an important role in the $\alpha$-clusterization; they are 
responsible for the formation of two $\alpha$-cluster state in $^{8}$Be 
\cite{OFE.06,EKNV.14} and linear chain of three $\alpha$-particles in $^{12}$C 
\cite{AJ.94,ZIM.15}.  

 The density distributions of other single-particle orbitals are characterized 
by different nodal structure. Their wavefunctions have a single node in
the direction perpendicular to the axis of symmetry, which in ideal case of
no state mixing would lead to zero density at the axis of symmetry. The [101]3/2 
and [101]1/2 orbitals show very similar density distributions of doughnut type 
in which the maximum of density is located in the equatorial plane 
[Figs.\ (\ref{Ar36-sp-p1})j-l and Figs.\ (\ref{Ar36-sp-p2})a-c].
These two orbitals at spin zero differ only in the 
orientation of the single-particle spin along the symmetry axis which has only 
moderate impact on the density distribution. At no rotation, these doughnut 
density distributions are axially symmetric. However, the rotation leads to 
a different redistribution of the neutron matter for the $r=\pm i$ branches of 
the single-particle orbital resulting in an asymmetric doughnut density distributions
in which the density depends on asimuthal angle.
For example, the matter is moved away from the $xz$ plane in the $\pm y$ directions for 
the $[101]1/2(r=-i)$ orbital [see Figs.\ (\ref{Ar36-sp-p2})a-c]. For the $[101]1/2(r=+i)$, 
this transition proceeds from the $yz$ plane in the $\pm x$ direction (similar to 
what is seen for the $[101]3/2(r=-i)$ orbital in Fig.\ (\ref{Ar36-sp-p1})j-l). 

 The wavefunction of the [211]3/2 orbital has one radial node and one node in the 
$z$-direction. As a result, its density distribution is the combination of two asymmetric 
density rings located symmetrically with respect of equatorial plane [see Figs.\ 
(\ref{Ar36-sp-p2})m-o]. The [211]1/2 orbital has similar structure with two density 
rings but in addition it has a spheroidal density cluster in the center of the nucleus 
(see Figs.\ \ref{Ar36-sp-p2}m-o). 
Three asymmetric density rings are seen in the [321]3/2 orbital (see Figs.\ 
\ref{Ar36-sp-p2}j-l). This asymmetry (dependence of the density on asimuthal 
angle) is due to rotation of the system; density rings in the  [211]3/2, 
[211]1/2 and [321]3/2 orbitals are axially symmetric at no rotation.

  The observed features of the single-particle density distributions coming from
the nodal structure of the wavefunction allow to understand in a relatively simple
way the necessary conditions for the $\alpha$-clusterization and for the formation 
of  nuclear molecules and ellipsoidal mean field type shapes. Two factors play an 
important role here: the degree of the localization of the wavefunction and the 
type of the density clusters formed by the single-particle orbital. It is clear 
that for the $\alpha$-clusterization the single-particle density clusters should be 
compact (well localized), should have spheroidal density distribution and overlap 
in space. These conditions are satisfied only for the lowest states of the $[NN0]1/2$ 
type with $N=0, 1$ and 2 which are active in the $\alpha$-cluster structures of very 
light nuclei \cite{OFE.06,EKNV.12,ZIM.15}.  With increasing particle number 
the orbitals with doughnut and multiply ring type density distributions 
become occupied. These states are substantially less localized; the maximum of the density in such 
structures is typically much smaller than the maximum of the density in the lowest $[NN0]1/2$
orbitals. In addition, such density distributions (doughnuts and rings) are
incompatible with $\alpha$-clusters. Thus, dependent on the nucleonic configuration 
they contribute into the building of either mean field structures or nuclear 
molecules. To build the later structures one has to move the matter from the 
neck (equatorial) region into the polar regions of the nucleus. Specific 
particle-hole excitation removing particles from (preferentially) doughnut type 
orbitals or from the orbitals which have a density ring in an equatorial plane 
into the orbitals (preferentially of the $[NN0]1/2$ type) which build the density 
mostly in the  polar regions will lead to more pronounced nuclear molecules. This
is what exactly happens in $^{36}$Ar on the transition from the hyperdeformed 
[4,4] configuration, which has ellipsoidal mean field like density distribution 
[see Fig. 24b in Ref.\ \cite{RA.16}]),  to the MD [31,31] configuration which
is an example of nuclear molecule [see Fig. 24c in Ref.\ \cite{RA.16} and 
Fig.\ \ref{Total-densities} in the present paper]. This transition involves the 
proton and neutron particle-hole excitations from the 3/2[321] orbital into the 
[440]1/2 orbital.

\section{Conclusions}
\label{conclusions}

  Nuclear shapes of two kinds at the ground state and in rotating
nuclei have been studied within the covariant density functional 
theory. 

Octupole shapes at the ground state have been searched in 
actinides and superheavy nuclei. The presence of the new region 
of octupole deformation in neutron-rich actinides with the center 
around $Z\sim 96, N\sim 196$ suggested in Ref.\ \cite{AAR.16} has 
been confirmed. However, our calculations do not predict octupole 
shapes in superheavy $Z\geq 108$  nuclei. The similarities and 
differences in the predictions of octupole deformation between 
non-relativistic and relativistic DFTs have been discussed.

  The role of the nodal structure of the wavefunction of
occupied single-particle orbitals in extremely deformed structures 
of the $N\sim Z$ nuclei has been investigated in detail on the 
example of megadeformed configuration in $^{36}$Ar. It allows to 
understand the formation of the $\alpha$-clusters in very light 
nuclei, the suppression of the $\alpha$-clusterization with the 
increase of mass number, the formation of ellipsoidal mean-field 
type structures and nuclear molecules. The particle-hole excitations 
between different types of the single-particle orbitals explain 
the transition between the later two classes of nuclear shapes.

\section{Acknowledgements}

 This material is based upon work supported by the U.S. Department 
of Energy, Office of Science, Office of Nuclear Physics under Award
No. DE-SC0013037.

\section{References}

\end{document}